\documentclass[prd, twocolumn, nofootinbib, floatfix]{revtex4-1}

\usepackage{amsmath}
\usepackage{graphicx}
\usepackage{dcolumn}
\usepackage{bm}
\usepackage{epsfig}
\usepackage{amssymb,latexsym,mathrsfs}
\usepackage{graphicx}
\usepackage{color}
\usepackage{hyperref}

\hypersetup{
    colorlinks=true,
    linkcolor=red,
    citecolor=blue,
} 

\newcommand{\be}{\begin{equation}}
\newcommand{\ee}{\end{equation}}
\newcommand{\bs}{\begin{split}} 
\newcommand{\bea}{\begin{eqnarray}}
\newcommand{\eea}{\end{eqnarray}}

\newcommand{\lcdm}{$\Lambda$CDM}

\newcommand{\kmax}{k_{\rm max}}

\newcommand{\vs}{\nonumber\\}

\begin{document}

\title{Redshift Space Distortion Reconstruction} 
\author{Alberto Vallinotto$^1$ \& Eric V.\ Linder$^{1,2}$} 
\affiliation{$^1$Berkeley Center for Cosmological Physics, Space Sciences 
Lab, \& Berkeley Lab, 
University of California, Berkeley, CA 94720, USA\\ 
$^2$Institute for the Early Universe WCU, Ewha Womans University, 
Seoul 120-750, Korea}

\begin{abstract}
Redshift space distortions of the matter density power spectrum carry 
information on the growth rate of cosmic structure but require accurate 
modeling of nonlinear and velocity effects on the density field.  We test 
and advance the reconstruction function of Kwan, Lewis, Linder (2012) in 
several ways.  First, we compare to the distribution function perturbative 
approach of Seljak \& McDonald (2012), deriving the mapping between them 
and showing how the KLL form can extend the validity of the latter.  
Second, using cosmological 
simulations rather than perturbation theory we calibrate the free functions 
within the KLL reconstruction form and fit their 
behavior in redshift and wavenumber.  
An efficient, new analysis pipeline rapidly calculates from simulation 
outputs the redshift space power spectrum.  
The end result is a robust analytic reconstruction function mapping the 
linear, real space matter power spectrum to the nonlinear, anisotropic 
redshift space power spectrum to an accuracy of $\sim$2-5\% in the range 
$k=0.1-0.5\,h$/Mpc for $z=0-2$.  As a by-product we derive RealEasy, 
an analytic mapping to the nonlinear, real space matter power spectrum 
(i.e.\ an analog to Halofit) to $\sim1\%$ accuracy over the same range.  
We also investigate an approach converse to KLL and find it to be 
complementary and capable of achieving an accuracy of $\sim$2-4\%  
over the range $k=0.3-1\,h$/Mpc for $z=0-2$. 

\end{abstract}

\date{\today} 

\maketitle

\section{Introduction} 

Three dimensional maps of galaxy clustering in large volumes of our 
Universe are becoming viable with the current and next generation of 
spectroscopic surveys (e.g.\ BOSS \cite{boss}, DESI \cite{desi}, 
PFS \cite{pfs}).  
These carry abundant, detailed information on 
the growth of large scale structure and the cosmological and astrophysical 
parameters affecting that structure.  In particular, since observations 
deliver information in redshift space, the data is a function of both the 
matter density and velocity field.  Nonlinear clustering amplifies the linear 
power on small scales, while velocities induced by mass 
inhomogeneities cause redshift space distortions that turn the real space 
power spectrum anisotropic. 

Thus observations are of the nonlinear, anisotropic redshift space power 
spectrum though theory is most adept at relating cosmological parameters to 
the linear, isotropic real space power spectrum.  In order to extract the 
cosmological information from the data we require robust maps between them. 
Many methods have been devised for carrying out this transformation 
\cite{kaiser,matsza,scocci,taruya,tang,reidwhite,jennings}, 
most relying on extending perturbation theory beyond the linear density 
regime.  Here we explore two more general prescriptions: the reconstruction 
function approach of \cite{kll} and the distribution function method of 
\cite{smd1}. 

The reconstruction function approach fits an analytic corrector function to 
the ratio between computational simulation results for the fully nonlinear 
redshift space density power spectrum and the linear 
real space density power spectrum, carrying out the mapping all in one 
step.  The derived reconstruction form, hereafter called KLL, can be 
physically motivated as combinations of nonlinear and velocity effects. 
It has been shown to be highly accurate, even out to wavenumbers 
$k=1\,h$/Mpc for dark matter and $k=0.6\,h$/Mpc for halos 
\cite{kll,julithesis}.  The KLL form has a fixed angular 
dependence but relies on simulations to determine the wavenumber $k$ and 
redshift $z$ dependence of its three free coefficients. 

The distribution function approach uses perturbation theory to expand 
the redshift space power spectrum in moments of the phase space distribution. 
Its form, 
hereafter called SMD, builds up from the linear theory in a perturbation 
sum and so is highly predictive but its accuracy breaks down as the 
wavenumber quantity used in the expansion moves outside its realm of 
validity. 

Here we explore the relation of these two approaches, combining the best 
features of each, and how simulations can further improve the reconstruction. 
In Sec.~\ref{sec:kllsmd} we derive the 
mapping from one approach to the other, compare the results of each 
to simulations, and consider their combination.  Section~\ref{sec:lcdm} 
uses simulations rather than perturbation theory to calibrate and investigate 
the redshift and wavenumber dependence of the KLL coefficients, and test 
the accuracy of its reconstruction.  Using these results we give an analytic 
fit to the real space power spectrum in Sec.~\ref{sec:real} and discuss an 
alternate redshift space approach in Sec.~\ref{sec:kfn}.  
We summarize and conclude in Sec.~\ref{sec:concl}.

\section{Mapping between KLL Reconstruction and SMD Perturbation Theory} \label{sec:kllsmd} 

Nonlinear structure formation maps a linear power spectrum at some 
redshift $z$ to a nonlinear power spectrum at $z$.  Redshift space 
distortions add a contribution due to the velocities along the line of 
sight, causing anisotropy with respect to the wavenumbers parallel to 
the line of sight, $k_\parallel=k\mu$, and perpendicular to the line 
of sight, $k_\perp=k\sqrt{1-\mu^2}$, where $\mu$ is the cosine of the 
wavevector with respect to the line of sight.  

\subsection{KLL reconstruction function} \label{sec:kll}

The KLL approach uses a reconstruction function $F(k,\mu,z)$ operating 
on some model for the power spectrum to give the full nonlinear, anisotropic 
redshift space power spectrum, 
\be 
P(k,\mu,z)=F(k,\mu,z)\,P_{\rm model}(k,\mu,z)\ . 
\ee  
The quantity $P(k,\mu,z)$ is sometimes written in the literature as $P^s$. 
The KLL form fitted to simulations in \cite{kll} is 
\be 
F(k,\mu,z)=\frac{A(k,z)}{1+B(k,z)\,k^2\mu^2}+C(k,z)\,k^2\mu^2 \ ,  
\label{eq:def_F}
\ee 
where the coefficients depend on the model to be corrected.  For example 
one could take $P_{\rm model}$ to be the linear real space power spectrum 
or some more complex prediction from higher order perturbation theory.  

We choose as $P_{\rm model}$ 
the linear theory redshift space power spectrum, i.e.\ the Kaiser formula 
\be 
P_{\rm Kaiser}(k,\mu,z)=[b(z)+f(z)\mu^2]^2\,P_L(k,z) \ , 
\ee 
where $P_L$ is the linear real space power spectrum, $f$ is the linear 
growth rate, and $b$ is the tracer (e.g.\ galaxy) bias.  While any model 
can be corrected accurately (see, e.g., \cite{kll} for examples using the 
models of \cite{scocci,taruya}), the Kaiser formula is correct in the limit 
of small wavenumber and is simply calculated.  Rather than doing a partial 
correction through higher order perturbation theory and another correction 
through $F$, we use the KLL form to account for the entire reconstruction. 

\subsection{SMD distribution function} \label{sec:smd}

Another approach, defined in \cite{smd1} and investigated further in 
\cite{smd2,smd3,smd4}, begins by considering the distribution function of 
particles in phase space $f(\vec{x}, \vec{q}, t)$. The dynamics of the 
system is controlled by the Vlasov-Poisson equation describing the evolution 
of $f(\vec{x}, \vec{q}, t)$. Since particles in a simulation are supposed 
to sample $f(\vec{x}, \vec{q}, t)$, this approach is appealing both 
theoretically and practically, as it can be translated straightforwardly 
into prescriptions to carry out measurements on N-body 
simulations (see \cite{smd2}). 

This formalism is based on introducing the following $n$-rank tensors 
\begin{eqnarray}
T^{n}_{i_1,i_2,...,i_n}(\vec{x})&\equiv&\frac{m}{\bar{\rho}}\int d^3\vec{q}\,f(\vec{x},\vec{q})\,u_{i_1}\,u_{i_2}\,...u_{i_n}\\
T^{n}_{i_1,i_2,...,i_n}(\vec{k}) &=& \int d^3\vec{x}\,e^{i\vec{k}\cdot\vec{x}}T^{n}_{i_1,i_2,...,i_n}(\vec{x}) \ , 
\end{eqnarray} 
where $\vec{q}=m\vec{u}$ is the comoving momentum. The density field in 
redshift space is then written as
\begin{eqnarray}
\delta_s(\vec{k})=\sum_n\frac{1}{n!}\left(\frac{ik\mu}{aH}\right)^n\,T_{\parallel}^{n}(\vec{k}) \ .
\end{eqnarray}
By defining $P^{ab}(\vec{k})\delta_D(\vec{k}-\vec{k'})\equiv\langle T_{\parallel}^a(\vec{k}) T_{\parallel}^{*b}(\vec{k}) \rangle$ the redshift space 
power spectrum gets expressed as 
\begin{eqnarray}
P(\vec{k})&=&\sum_{n=0}^{\infty}\frac{1}{(n!)^2}\left(\frac{k\mu}{aH}\right)^{2n}\,P^{nn}(\vec{k})\vs
&+&2 \, {\rm Re}\left[\sum_{a=0}^{\infty}\sum_{b>a}^{\infty}\frac{(-1)^b}{a!b!}\left(\frac{ik\mu}{aH}\right)^{a+b}P^{ab}(\vec{k})\right].\label{Eq:P_Uros}
\end{eqnarray}

At the same time, the angular dependence of the $T_{\parallel}^n(\vec{k})$ 
tensors can be expressed using spherical harmonics, 
\begin{eqnarray}
T_{\parallel}^n(\vec{k})=\sum_{l=n,n-2...}^{0\textrm{ or } 1}\sum_{m=-l}^{l}C_{lm}T_{lm}^{n}(k)Y_{lm}(\theta,\phi) \ , 
\end{eqnarray}
where $C_{lm}$ are normalization constants. Each $P^{ab}$ can then be 
expressed as a combination of these terms. Note that two sets 
of $l$'s and $m$'s will appear, belonging to the two $T_{\parallel}$'s 
that are being correlated.  However when averaging over the azimuthal 
angle $\phi$ all the terms with $m\neq m'$ will vanish. This then implies 
that 
\begin{eqnarray} 
P^{ab}(\vec{k})&=&\sum_{l=a,a-2,\dots}^{0\textrm{ or } 1}\ 
\sum_{l'=b,b-2,\dots}^{0\textrm{ or } 1} \label{Eq:Pabllm} \\ 
&\qquad&\sum_{m=0}^{{\rm min}(l,l')}\,P_{ll'm}^{ab}(k)
\mathcal{P}_{lm}(\mu)\,\mathcal{P}_{l'm}(\mu) \nonumber 
\end{eqnarray}
where $\mathcal{P}_{lm}$ denotes the associated Legendre 
polynomials.  

Thus the power spectrum is built up out of a momentum moment expansion in 
terms of 2-index power spectra $P^{ab}(\vec k)$, and these in turn are 
constructed from an angular expansion involving 5-index power spectra 
$P_{ll'm}^{ab}(k)$. Note that this formulation cleverly allows separation 
of the $k$ 
and $\mu$ dependence of the $P^{ab}$.  Once the $P_{ll'm}^{ab}(k)$ are 
computed or measured, all the rest follow. 

It is useful to write down a few of the $P^{ab}$ appearing in 
Eq.~(\ref{Eq:Pabllm}): 
\begin{eqnarray}
P^{00}(\vec{k})&=&P_{000}^{00}\\
P^{01}(\vec{k})&=&P_{010}^{01}\mu=-\frac{i\mu}{2k}\frac{dP^{00}(k)}{dt}\\
P^{11}(\vec{k})&=&P_{110}^{11}\mu^2+P_{111}^{11}(1-\mu^2)\vs
&=&\mu^2(P_{110}^{11}-P_{111}^{11})+P_{111}^{11}\\
P^{02}(\vec{k})&=&P^{02}_{000}-\frac{P^{02}_{020}}{2}+\frac{3\mu^2}{2}P^{02}_{020}\\
P^{12}(\vec{k})&=&P^{12}_{120}\left(\frac{3\mu^3-\mu}{2}\right)+P_{121}^{12}3\mu\left(1-\mu^2\right)\\
P^{22}(\vec{k})&=&\mu^4\left(\frac{9P_{220}^{22}}{4}-9P_{221}^{22}+9P_{222}^{22}\right)\vs
&+&\mu^2\left(9P_{221}^{22}-\frac{3P_{220}^{22}}{2}-18P_{222}^{22}+\frac{3P_{020}^{22}}{2}\right)\vs
&+&\frac{P_{220}^{22}}{4}+P_{000}^{22}-\frac{P_{020}^{22}}{2}+9P_{222}^{22}\\
P^{03}(\vec{k})&=&\mu \,P_{010}^{03}+\frac{P_{030}^{03}}{2}\left(5\mu^3-3\mu\right)\\
P^{13}(\vec{k})&=&P_{111}^{13}-\frac{3}{2}P_{131}^{13}+\mu^4\left(\frac{5}{2}P_{130}^{13}-\frac{15}{2}P_{131}^{13}\right)\vs
&+& \mu^2\left(P_{110}^{13}-P_{111}^{13}-\frac{3}{2}P_{130}^{13}+9P_{131}^{13}\right)\\
P^{04}(\vec{k})&=&P_{000}^{04}-\frac{P_{020}^{04}}{2}+\frac{3P_{040}^{04}}{8}+\frac{35\mu^4}{8}P_{040}^{04}\vs
&+&\mu^2\left(\frac{3}{2}P_{020}^{04}-\frac{30}{8}P_{040}^{04}\right)
\end{eqnarray} 

Now recall the definition of $P^{ab}_{ll'm}$ as a correlator of 
$T_{\parallel}$'s.  That is, 
\begin{eqnarray}
P^{ab}_{ll'm}(k)\, \delta(k-k')\equiv \langle T^a_{lm}(k)\,T^{*b}_{l'm'}(k')\rangle\delta_{mm'} \ . 
\end{eqnarray}
The physical meaning of the different terms contributing to the 5-index 
$P^{ab}_{ll'm}$ can be identified as correlations of the 
density/velocity/stress-energy (given by the index $a$) with 
density/velocity/stress-energy (given by the index $b$).

\subsection{Angular dependence comparison} 

The redshift space power spectra predicted by the KLL reconstruction function 
(times the Kaiser formula) can be directly compared to that from the SMD 
distribution function approach by identifying terms with common angular 
dependence, i.e.\ $\mu^0$, $\mu^2$, $\mu^4$, etc.  If one instead tries to 
compare terms with common $k$ dependence, one breaks the definition of $f$ 
as the linear growth rate and forces it to become $k$ dependent; furthermore 
the $k$ dependence of the $A$, $B$, $C$ coefficients is not known a priori -- 
this will be informed by the mapping.  

Within SMD, all $k$ dependence is separated from the angular $\mu$ dependence 
so one can write 
\begin{eqnarray}
P(k,\mu) &=& \sum_{n=0}^{\infty}\mu^{2n}\,\sum_{\alpha} S_{n\alpha}(k) =
\sum_{n=0}^{\infty}\mu^{2n}\, \bar{P}_{2n}(k) \,, 
\end{eqnarray} 
where $\alpha$ is a generic index that gathers together $l$, $l'$, etc., 
$S$ condenses the non-angular factors in Eq.~(\ref{Eq:P_Uros}), and 
$\bar{P}_{2n}$ further simplifies the sum.  
Of course since SMD relies on a perturbative expansion, this formalism will 
break down past some region of validity.  

\begin{figure*}[!htb]
\includegraphics[width=0.49\textwidth]{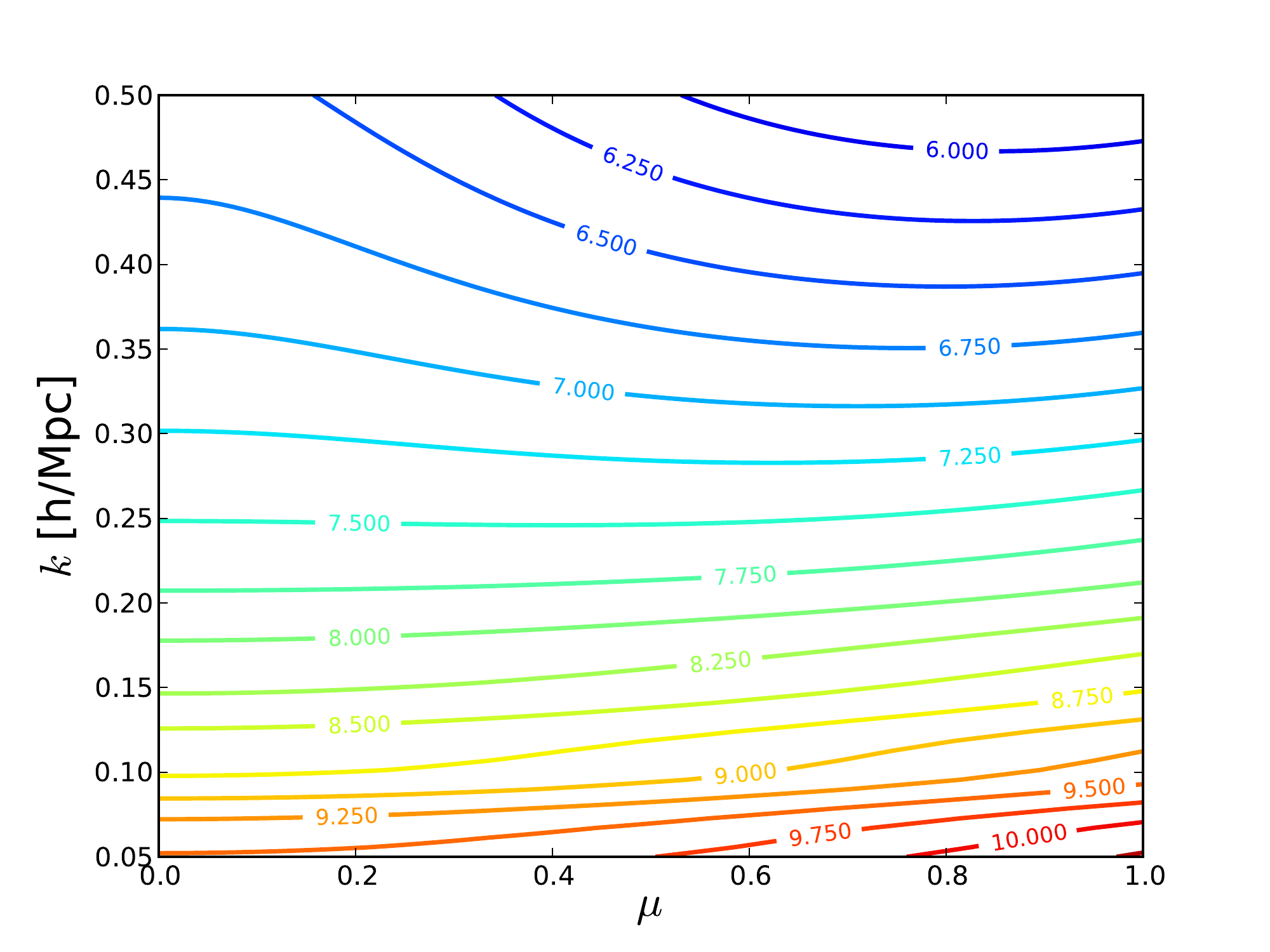}
\includegraphics[width=0.49\textwidth]{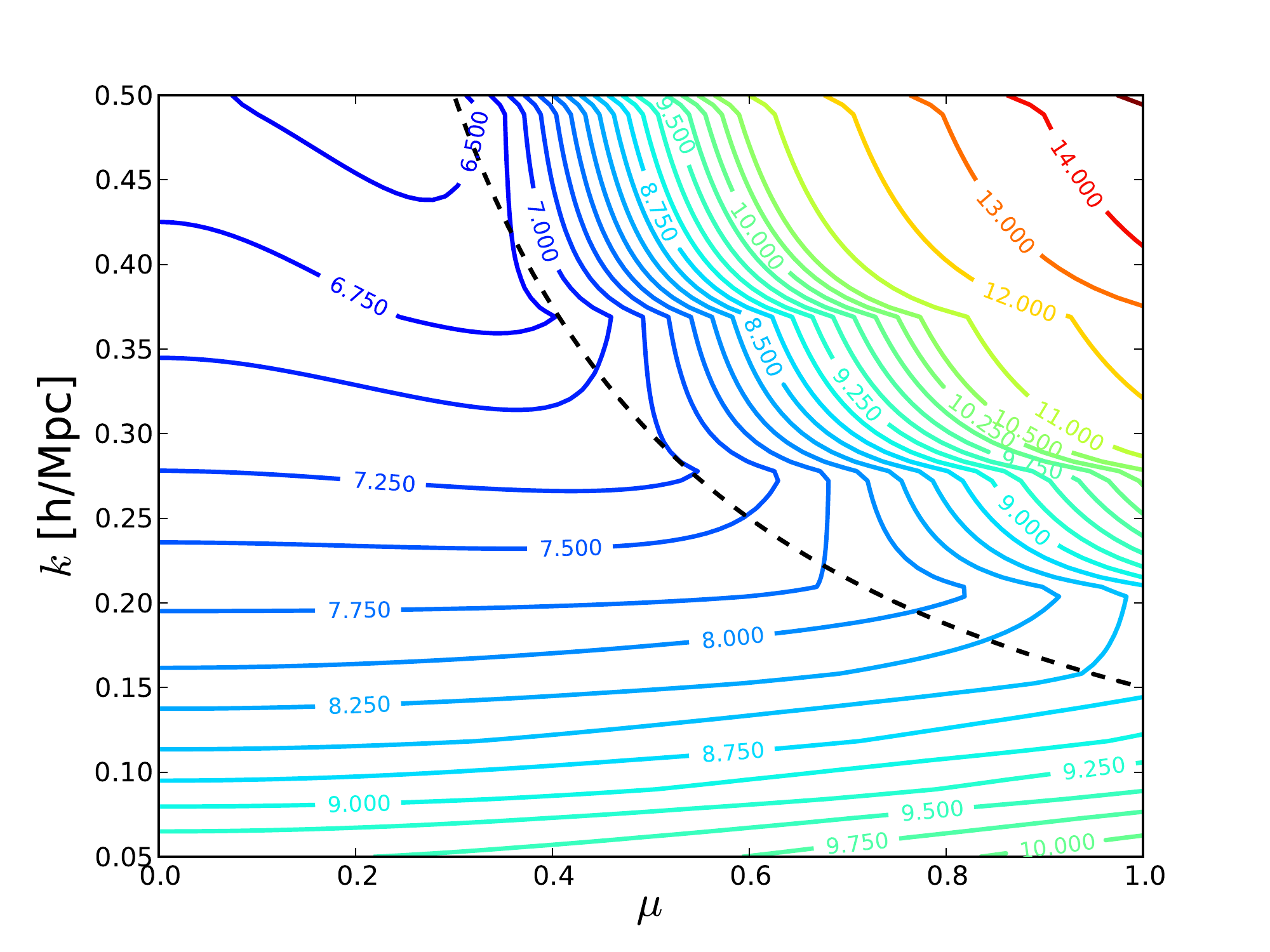}
\caption{Isocontours of the redshift space power spectra ($\ln P$) are shown 
for the KLL 
approach (left panel) and for the SMD approach (right panel).  The two 
methods agree for small $k\mu$ and KLL is known to match numerical 
simulations, but the SMD approach breaks down around $k\mu\gtrsim0.15\,h$/Mpc 
(given by the dashed curve). 
}
\label{fig:Ps} 
\end{figure*}

Figure~\ref{fig:Ps} plots the logarithm of the redshift space power spectra 
in the $k$--$\mu$ plane, for the KLL (left panel) and SMD (right panel) 
cases.  We work at redshift $z=0$ as the greatest challenge to accurate 
modeling. Three characteristics are apparent: 1) Good agreement in the 
linear region ($k\ll1$), 2) Good agreement at small angles to the line of 
sight ($\mu\ll1$), and 3) Breakdown of SMD beyond $k\mu\gtrsim0.15\,h$/Mpc. 
This limit to the SMD approach agrees with that reported 
in \cite{smd2}.  We emphasize that the KLL results shown come purely 
from the analytic prediction using the coefficients given in \cite{linsam}, 
with no adjustable parameters (\cite{linsam} roughly estimated the 
coefficients; we will calibrate them using a simulation analysis pipeline 
later in this paper).

\begin{figure*}[!htb]
\includegraphics[width=0.49\textwidth]{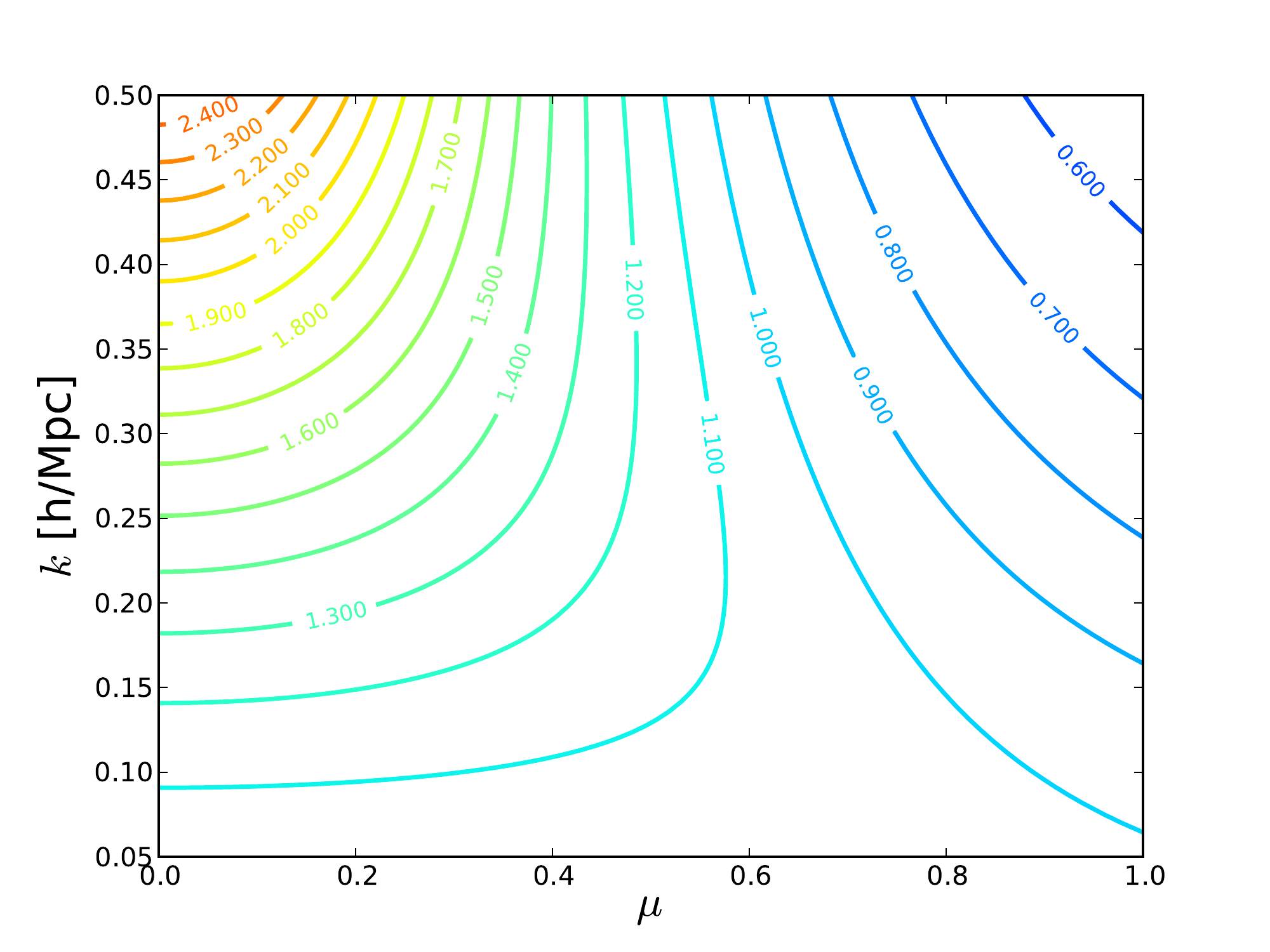}
\includegraphics[width=0.49\textwidth]{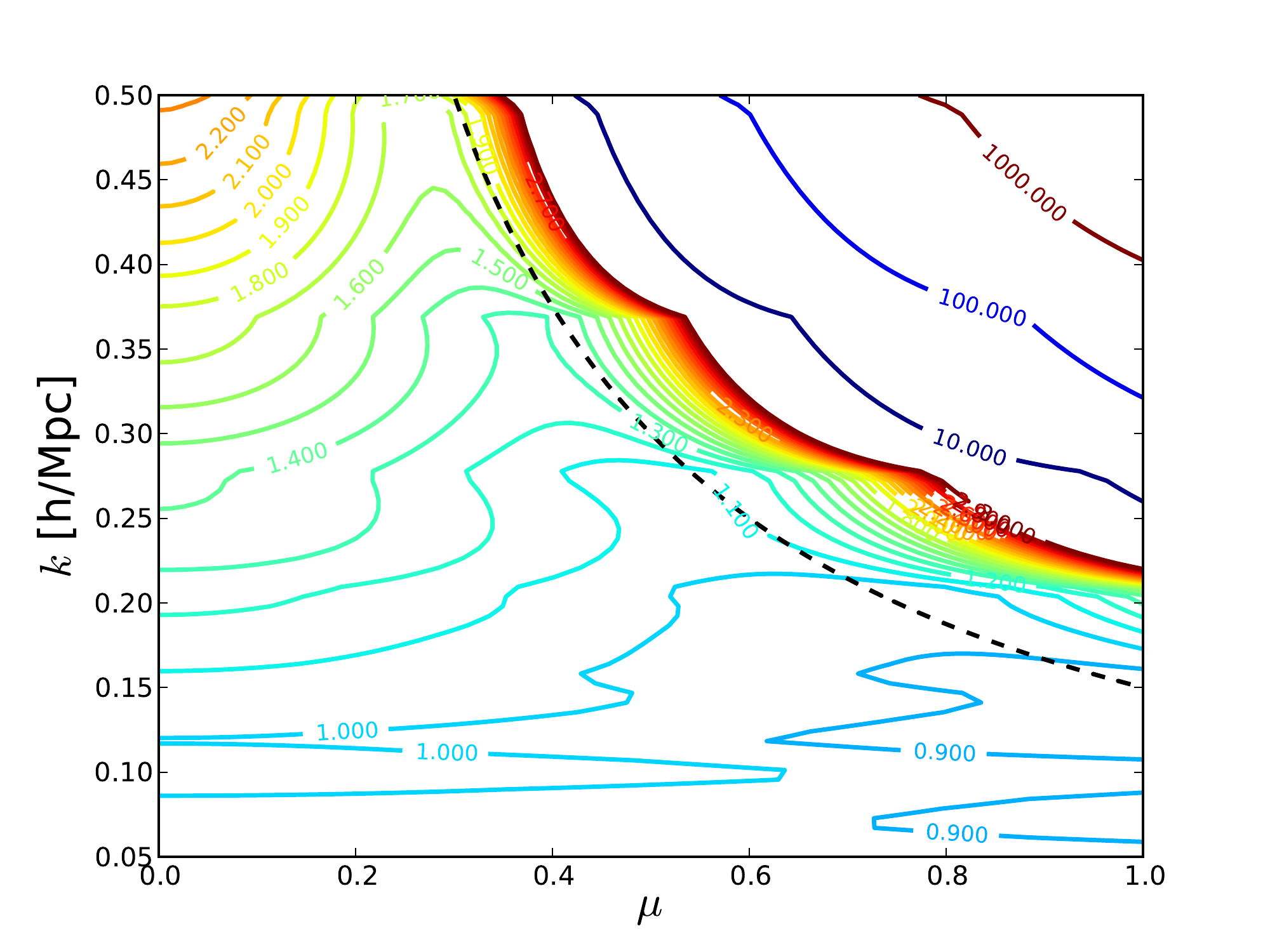}
\caption{\label{fig:F} Reconstruction (``corrector'') functions for the 
KLL (left) and SMD (right) formalisms are shown in the $k$--$\mu$ plane.} 
\end{figure*}

As an alternate view, we show in Fig.~\ref{fig:F} the reconstruction 
functions necessary to take the linear real space power spectrum to the 
predicted nonlinear redshift space power spectrum in each approach.  We 
see that this correction in the SMD case runs away to large values for 
$k\mu\gtrsim0.15\,h$/Mpc.

While keeping in mind the range of validity, we can now proceed to the 
comparison of KLL and SMD term by term in the $\mu^n$ expansion.  For the 
terms independent of $\mu$ (i.e.\ isotropic), equating the terms implies 
\be 
P_L(k)\,A(k) = \bar{P}_0(k) = P_{NL}(k) \ , \label{eq:p0} 
\ee 
where $P_{NL}$ is the nonlinear real space power spectrum, 
so $A$ is just the ratio of the nonlinear to linear isotropic 
power spectra, as also found by \cite{kll}. 

The next two terms, depending on $\mu^2$ and $\mu^4$, give 
\bea 
&&2f P_L\,A + K^2 P_L\,(C-AB) = \bar{P}_2 \label{eq:p2}\\ 
&&f^2 P_L + 2 K^2 f P_L\,(C-AB) + K^4 A B^2 P_L = \bar{P}_4 \label{eq:p4} 
\eea 
where $K\equiv k/(aH)$.  The quantities $B$ and $C$ are mixed together, 
involving the autocorrelations of the first moment of the phase space 
distribution, i.e.\ the velocity field, but also nonlinearities and 
cross-correlations in the density and the stress-energy (see also 
\cite{smd2,kll}).

\begin{figure}[!htb]
\includegraphics[width=\columnwidth]{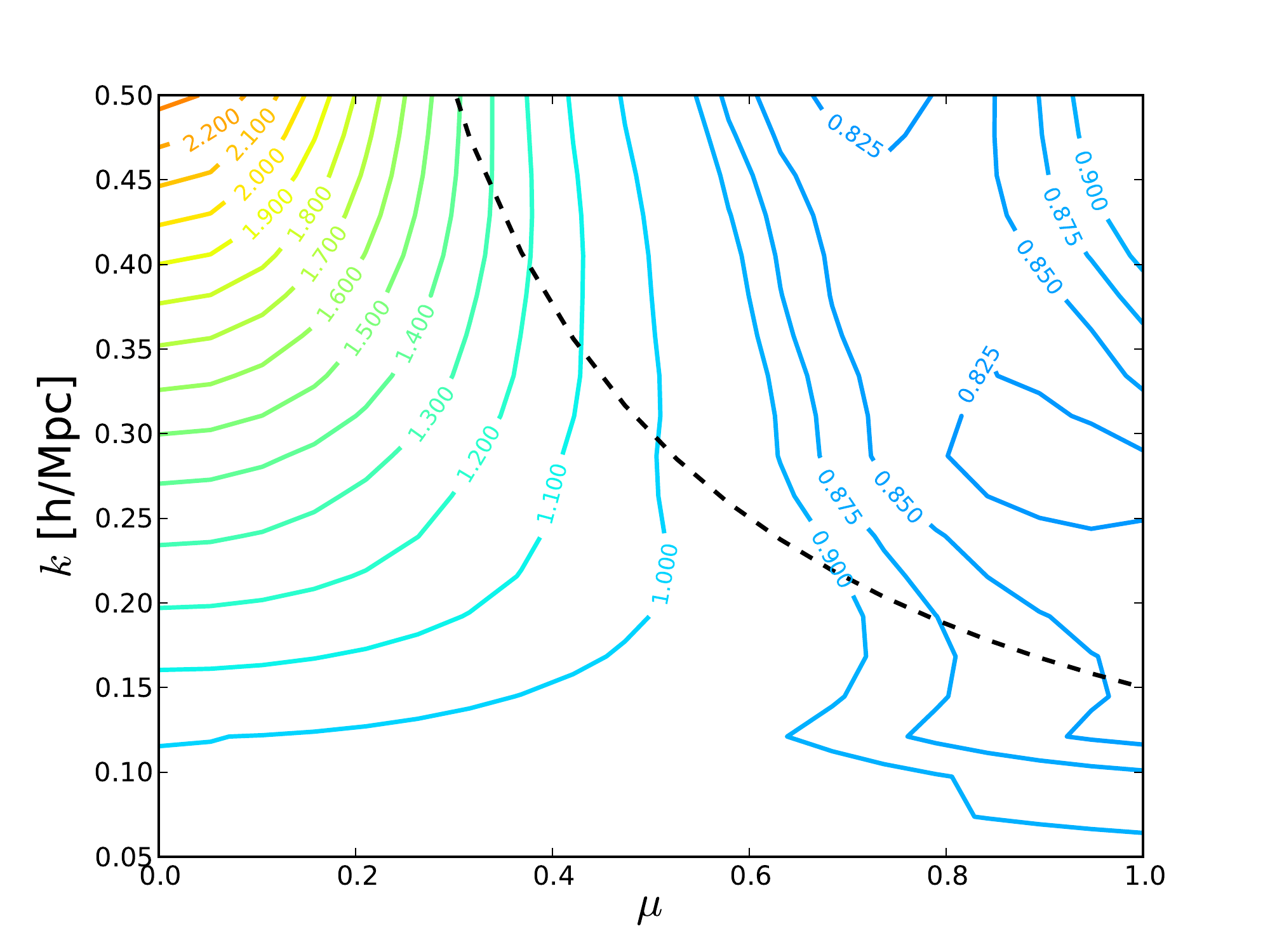}
\caption{The reconstruction function is shown 
for coefficients constructed using Eqs.~(\ref{eq:A}-\ref{eq:C}) 
and measurements from the simulations of \cite{smd2} but 
using the KLL form.  This 
successfully removes the dramatic breakdown seen in the right panel of 
Fig.~\ref{fig:F}. 
} 
\label{fig:FT} 
\end{figure}  

Solving these three equations for $A$, $B$, and $C$ 
yields the mapping between the two approaches: 
\begin{eqnarray}
A&=&\frac{P_{NL}}{P_L}\label{eq:A}\\
B^2&=&\frac{\bar{P}_4-2f\bar{P}_2}{K^4\,P_{NL}}+3\frac{f^2}{K^4}\\
C&=&AB+\frac{\bar{P}_2-2fP_{NL}}{K^2\,P_L} \ . \label{eq:C} 
\end{eqnarray} 
Thus $\bar P_0$, $\bar P_2$, $\bar P_4$ can be determined by $A$, $B$, and 
$C$ through Eqs.~(\ref{eq:p0})--(\ref{eq:p4}), or conversely $A$, $B$, and 
$C$ can be found by Eqs.~(\ref{eq:A})--(\ref{eq:C}).  Also note that 
the coefficients always enter in KLL as $BK^2$ and $CK^2$. 

At higher orders, since $A$, $B$, $C$ are already determined, the KLL 
terms will differ in general from the moment method expansion terms.  
(Note that KLL implicitly includes all orders of $\mu$.)  
However, the higher order SMD terms are negligible except at $k>0.1\,h$/Mpc 
\cite{smd2} and SMD quickly loses accuracy at higher $k$.  
So the mapping we have established is sufficient.  

Indeed we can use this mapping to merge the strengths of each approach: 
the physical basis at low $k\mu$ of the perturbation theory in SMD with 
the analytic reconstruction over a wide range of $k\mu$ in the KLL form. 
In the next section we apply this by {\it deriving\/} $A$, $B$, $C$ from 
the SMD 2-index power spectra of their simulations (rather than using the 
values in \cite{linsam}), then feeding these into the KLL reconstruction 
form and comparing 
to the full nonlinear redshift space power spectrum of the simulation. 
We will see this use of the KLL form significantly extends the range of 
accuracy of the reconstruction.

%
\begin{figure}[!htb]
\includegraphics[width=\columnwidth]{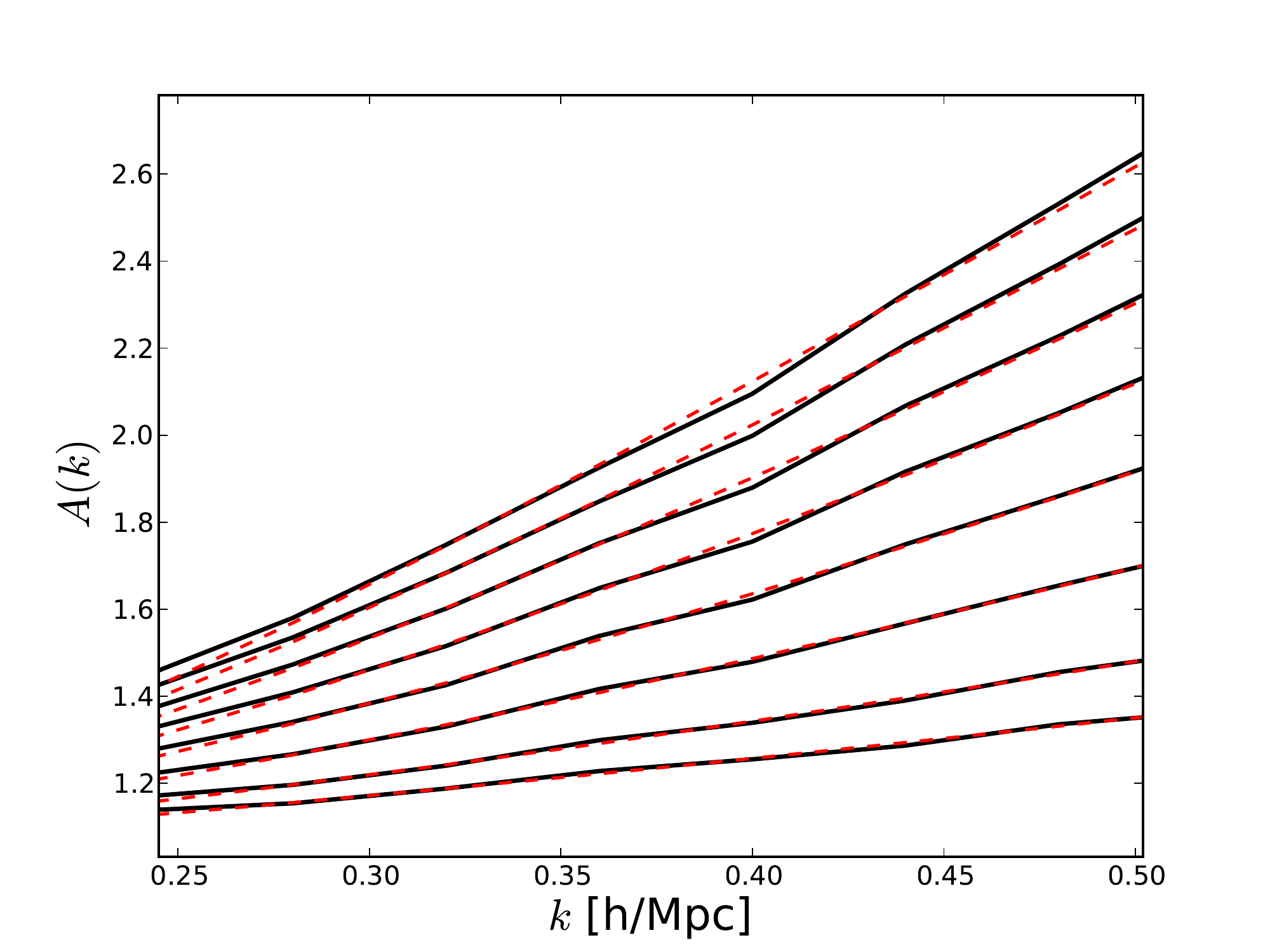}\\ 
\includegraphics[width=\columnwidth]{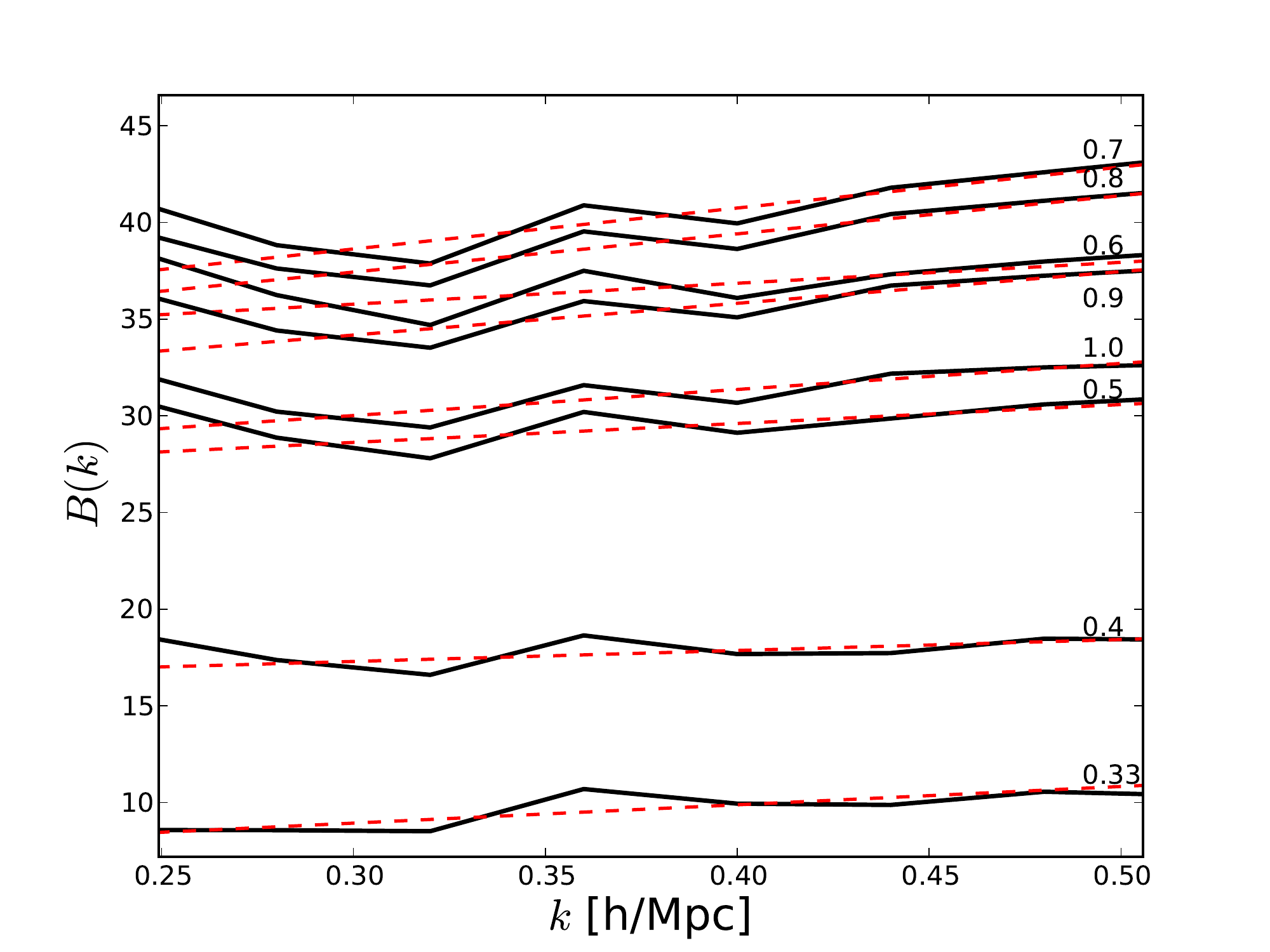}\\ 
\includegraphics[width=\columnwidth]{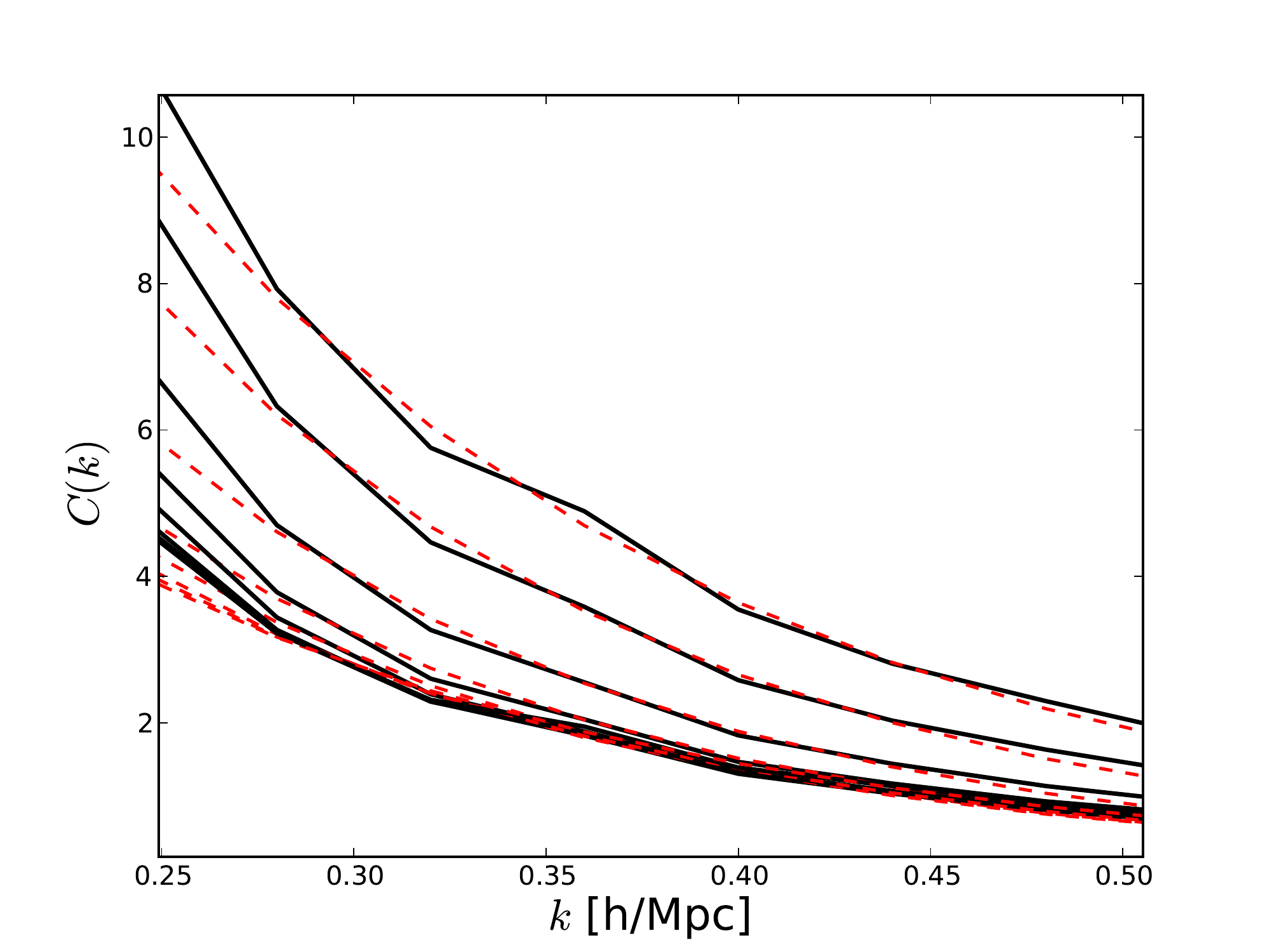}
\caption{KLL coefficient functions $A(k)$, $B(k)$, and $C(k)$ as measured 
from the M000 model of the Coyote Universe simulation suite are plotted 
for $z=\{0, 0.11, 0.25, 0.43, 0.67, 1, 1.5, 2\}$, from top to bottom for 
$A$, bottom to top for $C$, and as labeled for $B$.  The red, dashed lines 
are simple analytic fits. 
} 
\label{fig:ABC_M000} 
\end{figure}  
%

\subsection{Applying the mapping} \label{sec:compare} 

Summing the measurements of $P^{ab}$ (kindly provided to us by Teppei Okumura) 
on the simulations in \cite{smd2}, we can get the terms 
$\bar P_{2n}$.  From these, 
Eqs.~(\ref{eq:A})--(\ref{eq:C}) provide $A$, $B$, and $C$.  Those 
coefficients are then used within the usual KLL reconstruction function 
to transform the linear power spectrum $P_L$ (and the Kaiser formula) 
into a predicted nonlinear redshift space power spectrum. 

The derived reconstruction function is shown in Fig.~\ref{fig:FT}.  
Comparing to the right panel of Fig.~\ref{fig:F}, the original SMD approach, 
we see a marked improvement, with no sign of the cliff beyond 
$k\mu\gtrsim0.15\,h$/Mpc where the SMD function jumps upward by a factor of 
a thousand.  Rather, the reconstruction is now much smoother and in 
qualitative accord with the pattern of the left panel of Fig.~\ref{fig:F}, 
the baseline KLL approach.  There is no obvious breakdown 
in going to higher $k\mu$, one of the virtues of the KLL method's implicit 
inclusion of all orders of $\mu$. 

These results highlight the power of both the KLL form and the advantage 
of using simulations to estimate accurately the coefficients $A(k,z)$, 
$B(k,z)$, $C(k,z)$.  
In the next section we leave perturbation theory behind and 
develop an efficient computational pipeline 
to extract from simulations accurate fits for $A$, $B$, 
and $C$.  We apply this to a high definition realization of the Coyote 
cosmological simulation suite as an example.

\section{Calibrating KLL Coefficients with Simulations} \label{sec:lcdm} 

In order to calibrate the KLL coefficients from cosmological simulations 
we have developed a parallel MPI code that, starting from a simulation 
snapshot, computes the redshift space power spectrum and fits the 
reconstruction coefficients as a function of $k$ at each redshift.  
This automated analysis pipeline is flexible enough to handle large 
simulation boxes.  
As is often the case, the limiting factor in this analysis is the amount 
of memory per rank: since the FFT is carried out by decomposing the 
simulation box into slabs and assigning each slab to a rank, the largest 
number of cells into which the snapshot can be decomposed is determined by 
the amount of memory required for a single layer slab (i.e.\ a 3D slab of 
dimensions $N\times N \times 1$, where N is the number of cells the snapshot 
is decomposed into). 

The operations carried out on the data ranks have been 
decomposed into a set of simple ``atomic'' operations -- while this choice 
might seem an extra complication, this allows the implementation of other 
kinds of analyses using the same framework. In particular, the implementation 
of the measurement of the SMD power spectra $P^{ab}(\vec k)$ and of the 
divergence of the velocity field, if desired, become straightforward. 
Furthermore, the implementation of the particle deposition scheme is 
flexible enough to allow reconstructing the density field starting from 
``particles'' of different masses; within this framework it is therefore 
straightforward to carry out measurements also using halos (although we 
leave this for future work).

\begin{figure}[!htb]
\includegraphics[width=\columnwidth]{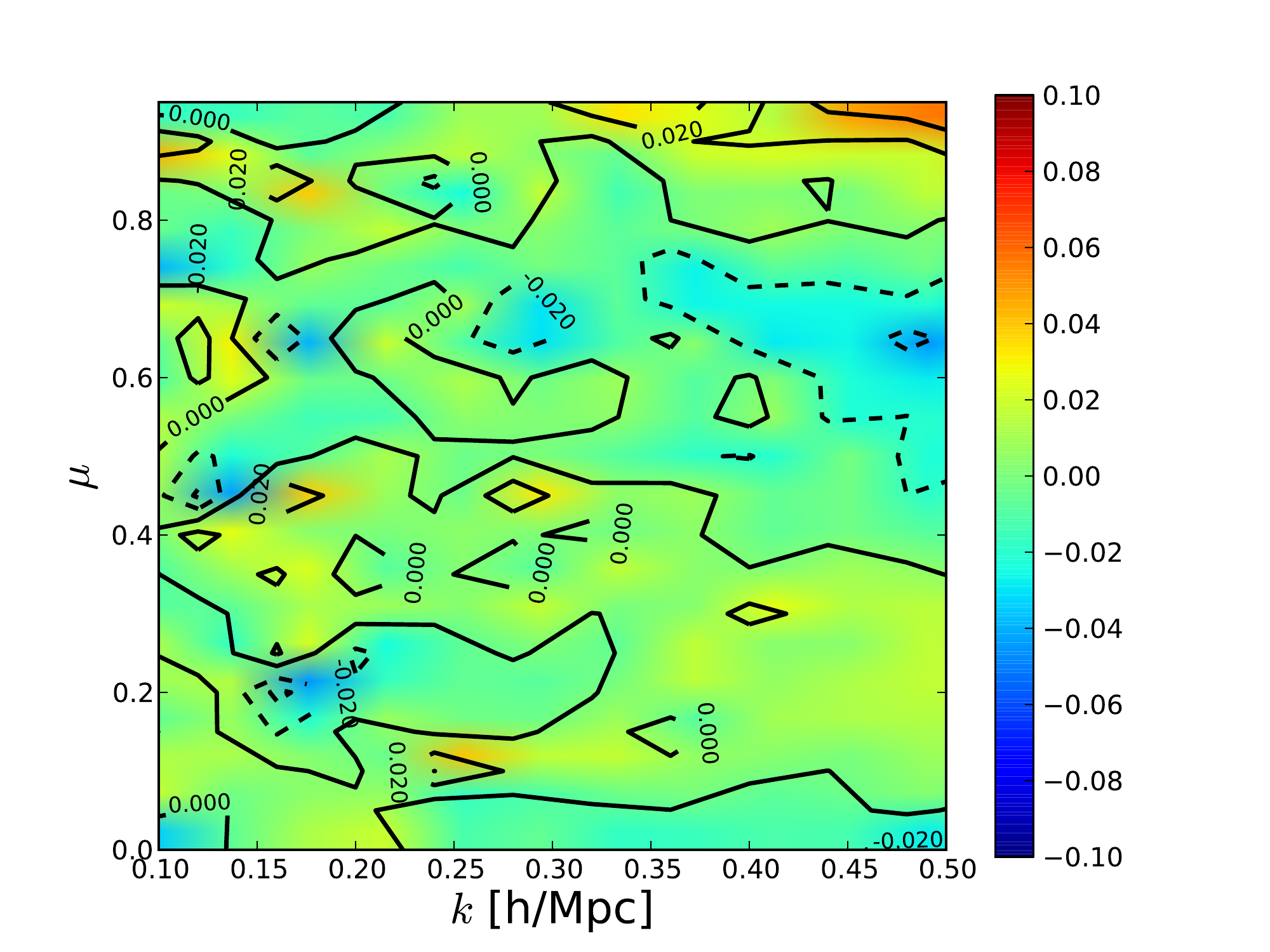}\\ 
\includegraphics[width=\columnwidth]{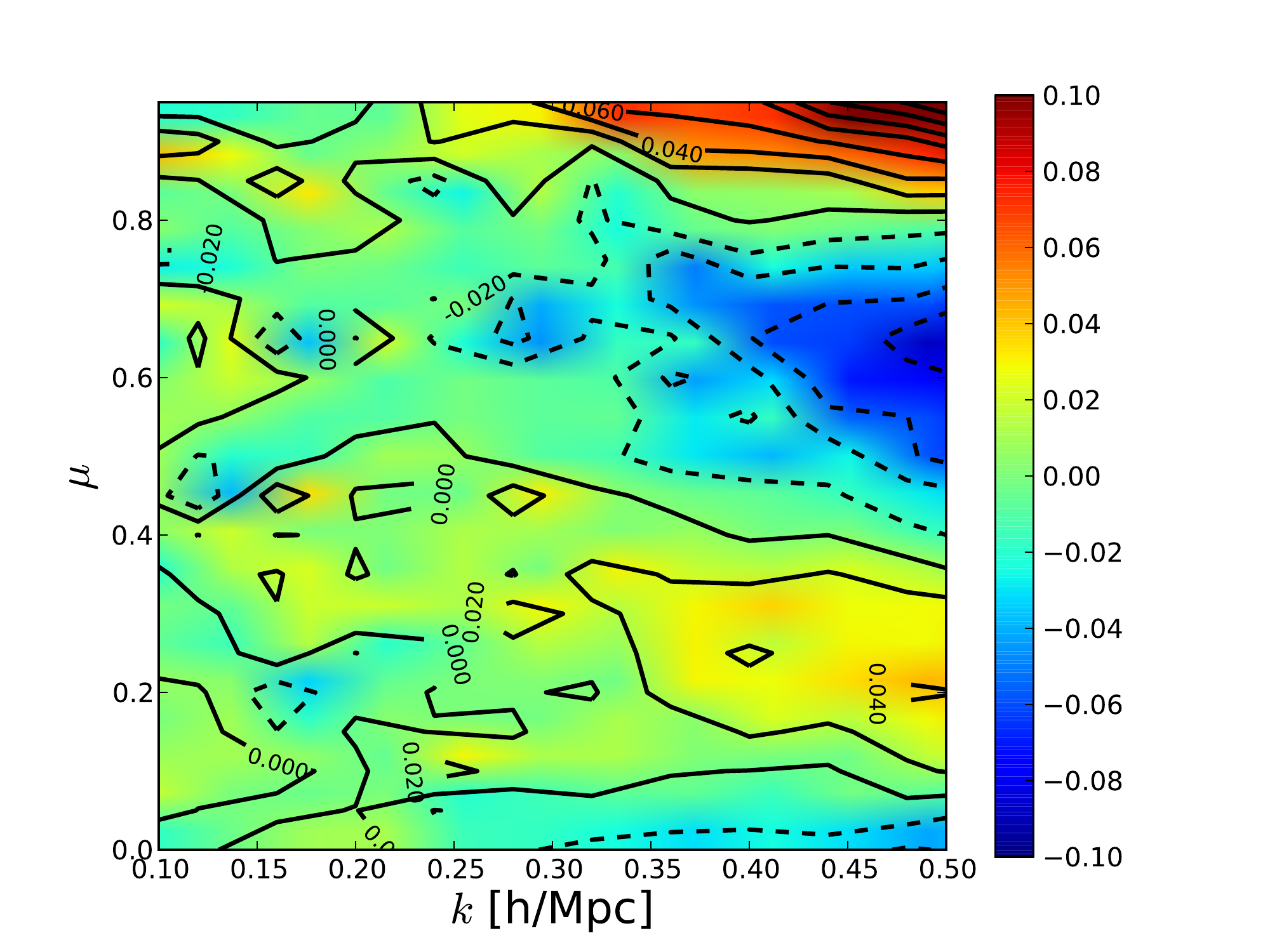}\\ 
\includegraphics[width=\columnwidth]{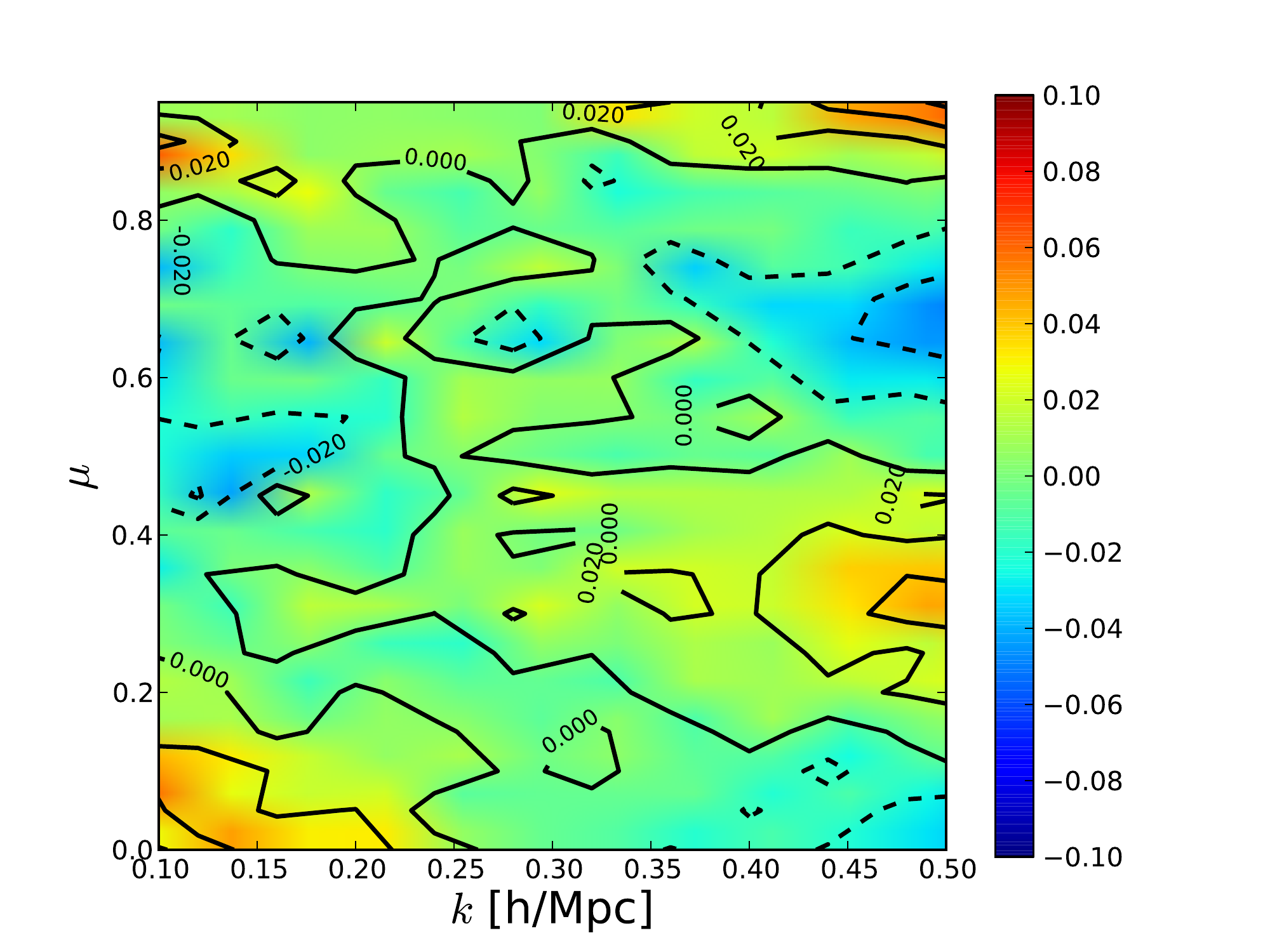} 
\caption{Fractional difference between the RSD power spectrum measured on 
M000 simulation and the one reconstructed using the best fit calibrated 
values of $A(k)$, $B(k)$, and $C(k)$, at $z=0$, 0.67, 2 (top, center, 
bottom panel). 
} 
\label{fig:RSD_PS_fractional_difference_M000} 
\end{figure}

Briefly, the steps taken to carry out the measurement and fit are:
\begin{enumerate}
\item Load the particles and, if calculating the redshift space distorted density field, displace them according to their velocity along a predetermined direction that has been chosen as the line of sight.
\item Reconstruct the density field $\delta(\vec{x})$ on a 3D grid using a 
cloud in cell deposition scheme.
\item Fourier transform the density field and deconvolve the cloud in cell deposition scheme to obtain $\delta(\vec{k})$.
\item Square the Fourier transformed density field.
\item Bin the results in both $\{k_{\perp},k_{\parallel}\}$ and $\{k,\mu\}$ and thus measure the power spectra.
\item Measure $F(k,\mu)$ from
\begin{equation}
F(k,\mu,z)=\frac{P(k,\mu,z)}{P_L(k,\mu,z)[1+f(z)\mu^2]^2},
\end{equation}
where for the linear power spectrum we use the one measured on the initial conditions,  multiplied by appropriate linear growth to redshift $z$.
\item For each bin in $k$-space $k_i$ and each redshift $z_j$ we then find the best fit value for $A(k_i, z_j)$, $B(k_i, z_j)$ and $C(k_i, z_j)$.
\end{enumerate}

As a first application, we use our analysis pipeline on the high resolution box of the M000 model of the Coyote Universe simulation suite \cite{Heitmann:2008eq,Heitmann:2009cu,Lawrence:2009uk}: a flat $\Lambda$CDM cosmology with 
dimensionless physical matter density $\omega_m=0.1296$, physical baryon 
density $\omega_b=0.0224$, scalar tilt $n_s=0.97$, amplitude of mass 
fluctuations today $\sigma_8=0.8$, and reduced Hubble constant today 
$h=0.72$ containing $1024^3$ dark matter particles in a box of 1300 Mpc 
evolved starting 
from $z=200$ using the publicly available code GADGET-2. 

Having access only to the high resolution realization of the Coyote Universe suite restricts the accuracy with which the power spectrum can be measured on large scales. The fractional precision is 
approximately given by the inverse square root of the number of measurable 
modes, thus
\bea 
\frac{\delta P}{P}&\approx& N_{\rm modes}^{-1/2}
=\left(\frac{3}{2}\frac{k^2 \Delta k \Delta\mu V_{\rm box}}{4\pi^2}\right)^{-1/2}\\
&\approx& 1.61\left(\frac{k}{0.25}\right)^{-1} \left(\frac{\Delta k}{0.04}\right)^{-1/2} \left(\frac{\Delta \mu}{0.05}\right)^{-1/2} \% \ 
\eea 
where the 3 in the $3/2$ reflects the 3 orthogonal directions through the 
box and the 2 comes from using the real part of the Fourier modes.  We have 
verified the results through actual mode counting. 

Currently, we restrict our analysis to the regime where the simulation 
scatter $\delta P/P\lesssim 2\%$, since that will be our reconstruction 
accuracy goal, corresponding to $k\gtrsim0.25\,h$/Mpc.  
We regard this investigation 
as a proof of principle, since given success here there should be no 
substantial obstacle to increasing its precision through multiple 
realizations, extending its range to larger scales (smaller $k$), and 
connecting it down to the perturbation theory region.  
On the high $k$ end (small scales), we will explore out to  
$\kmax=0.5\,h$/Mpc, although our reconstruction function works reasonably 
beyond this.  The analytic fits we discuss next are therefore applied to 
the range $k=0.25-0.5\,h$/Mpc.

We optimize the reconstruction function coefficients so as to minimize the residuals of the redshift space power spectrum relative to the simulation results over the $k$-$\mu$ space.  We adopt $\Delta k=0.04$ and $\Delta \mu=0.05$ and have also checked that the joint $k$-$\mu$ optimization is consistent with a $\mu$ optimization at each fixed $k$.

\begin{figure}[!htb]
\includegraphics[width=\columnwidth]{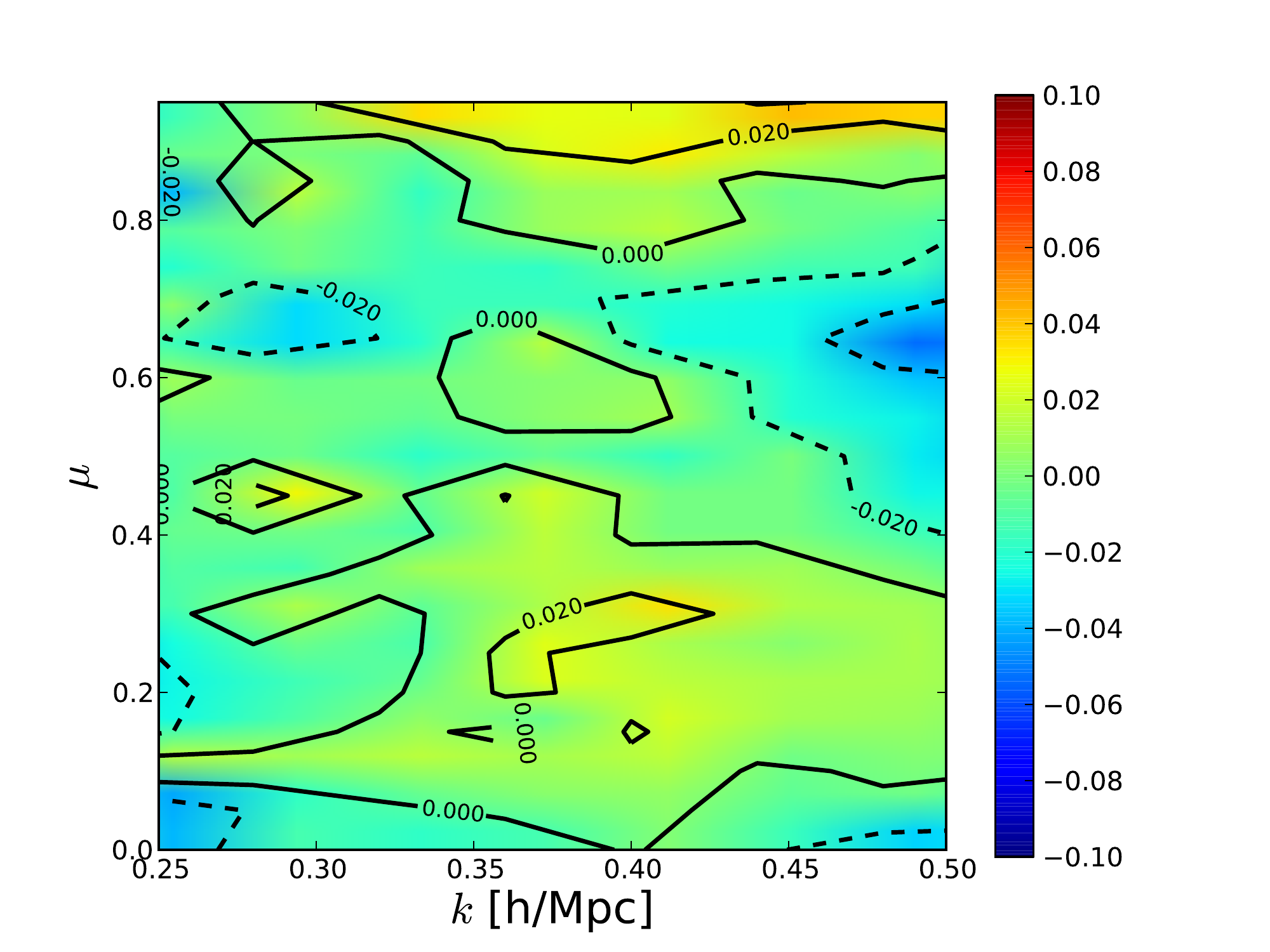}\\ 
\includegraphics[width=\columnwidth]{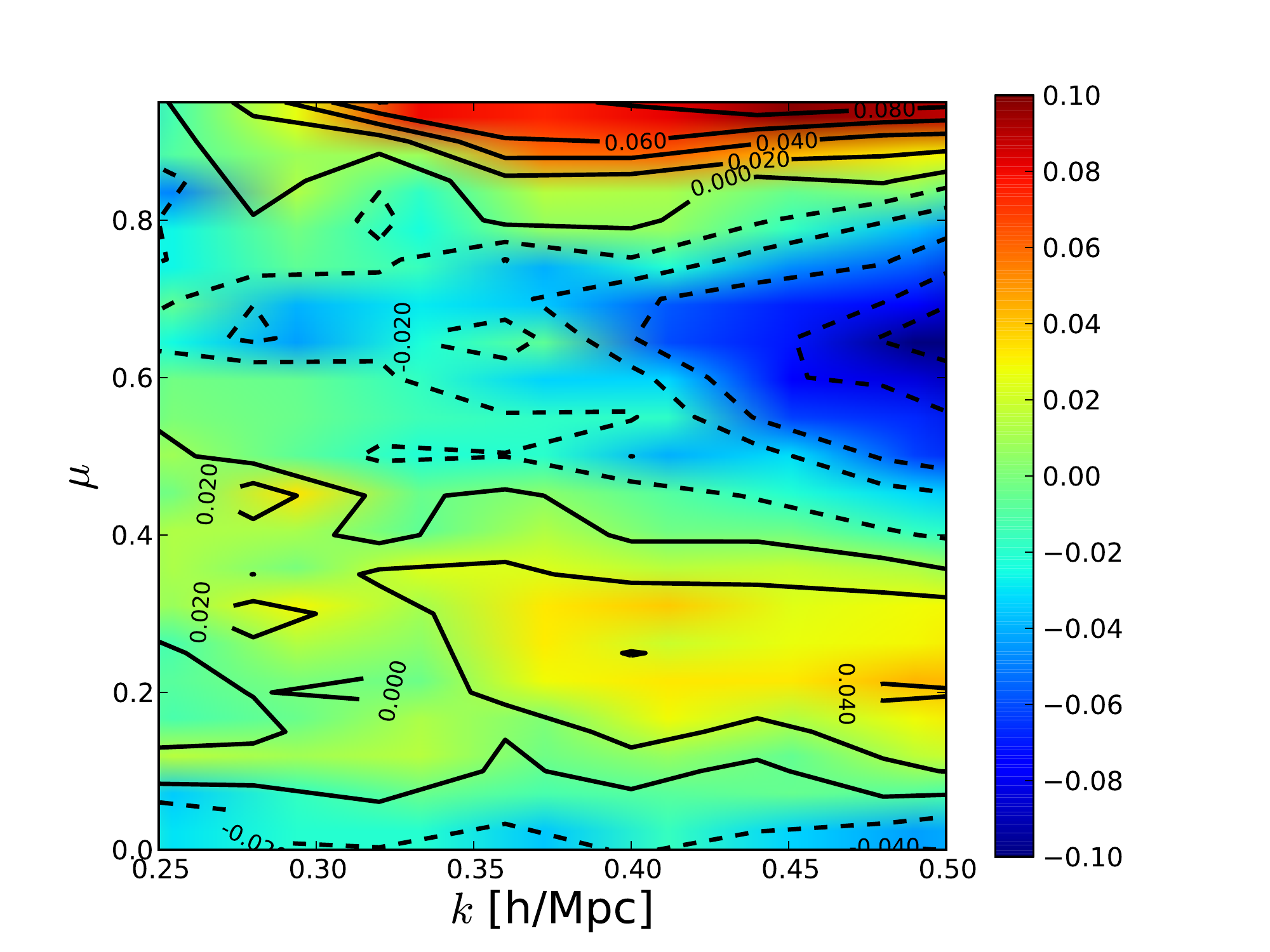}\\ 
\includegraphics[width=\columnwidth]{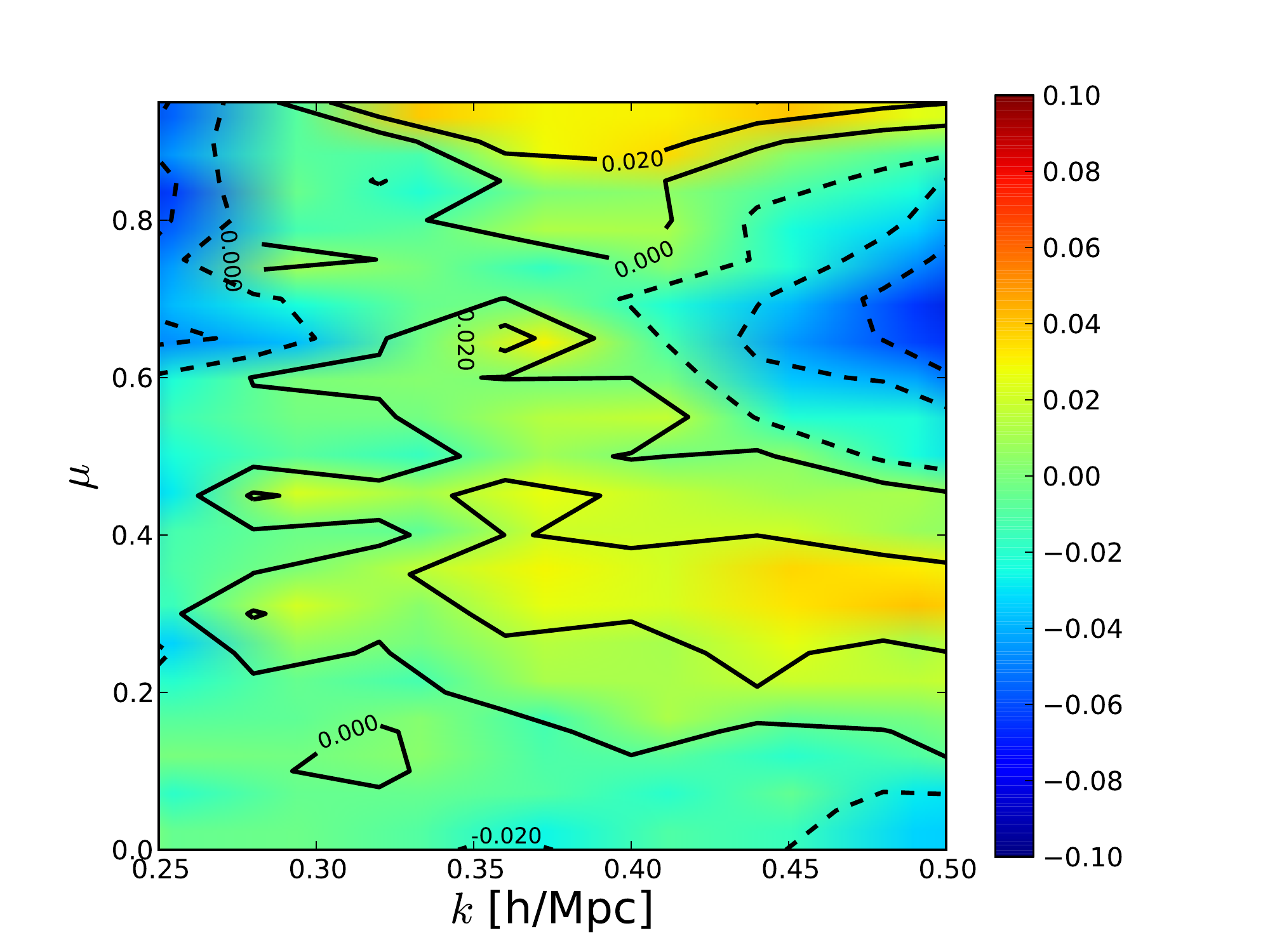} 
\caption{Fractional difference between the RSD power spectrum measured on 
M000 simulation and the one using the simple analytic reconstruction from 
Eqs.~(\ref{eq:anlya}), (\ref{eq:anlyb}), (\ref{eq:anlyc}), 
at $z=0$, 0.67, 2 (top, center, bottom panel). 
} 
\label{fig:Anly_PS_fractional_difference_M000} 
\end{figure}

The best fit functions $A(k)$, $B(k)$ and $C(k)$ obtained for 
$a=\{1,0.9,0.8,0.7,0.6,0.5,0.4,0.333\}$, i.e.\ 
$z=\{0, 0.11, 0.25, 0.43, 0.67, 1, 1.5, 2\}$ are shown in 
Fig.~\ref{fig:ABC_M000}.  
We find the reconstruction function coefficients are well behaved, with 
simple forms in $k$ and somewhat more complicated dependence on redshift. 
We discuss each coefficient in turn.  The deviation of the factor 
$A(k,z)$ from unity arises from nonlinearity and so as expected $A$ increases 
with $k$ and decreases with redshift (increases with $a$).  The quantity 
$B$ acts like a velocity dispersion, including nonlinear effects, and as 
expected is small at large redshift, increasing as nonlinearity does.  
However, the velocity effects also decline toward the present as the growth 
rate $f$ does as dark energy comes to dominate -- recall that the velocity 
damping often enters in the combination $f^2\sigma_v^2 k^2\mu^2$ 
(cf.\ $Bk^2\mu^2$ in KLL). 
These competing factors cause $B$ to be nonmonotonic with redshift, reaching 
a maximum amplitude around $z\approx0.6$.  The $k$ dependence of $B$, however, 
is small over this range, with a slight linear trend out to 
$k\approx0.8\,h$/Mpc.  
The coefficient $C$ also involves a mix of velocity and nonlinear effects, 
and acts to restore some power lost to the damping.  This quickly dies off 
(roughly exponentially) with $k$, however, and also with decreasing redshift.

\begin{figure}[!htb]
\includegraphics[width=\columnwidth]{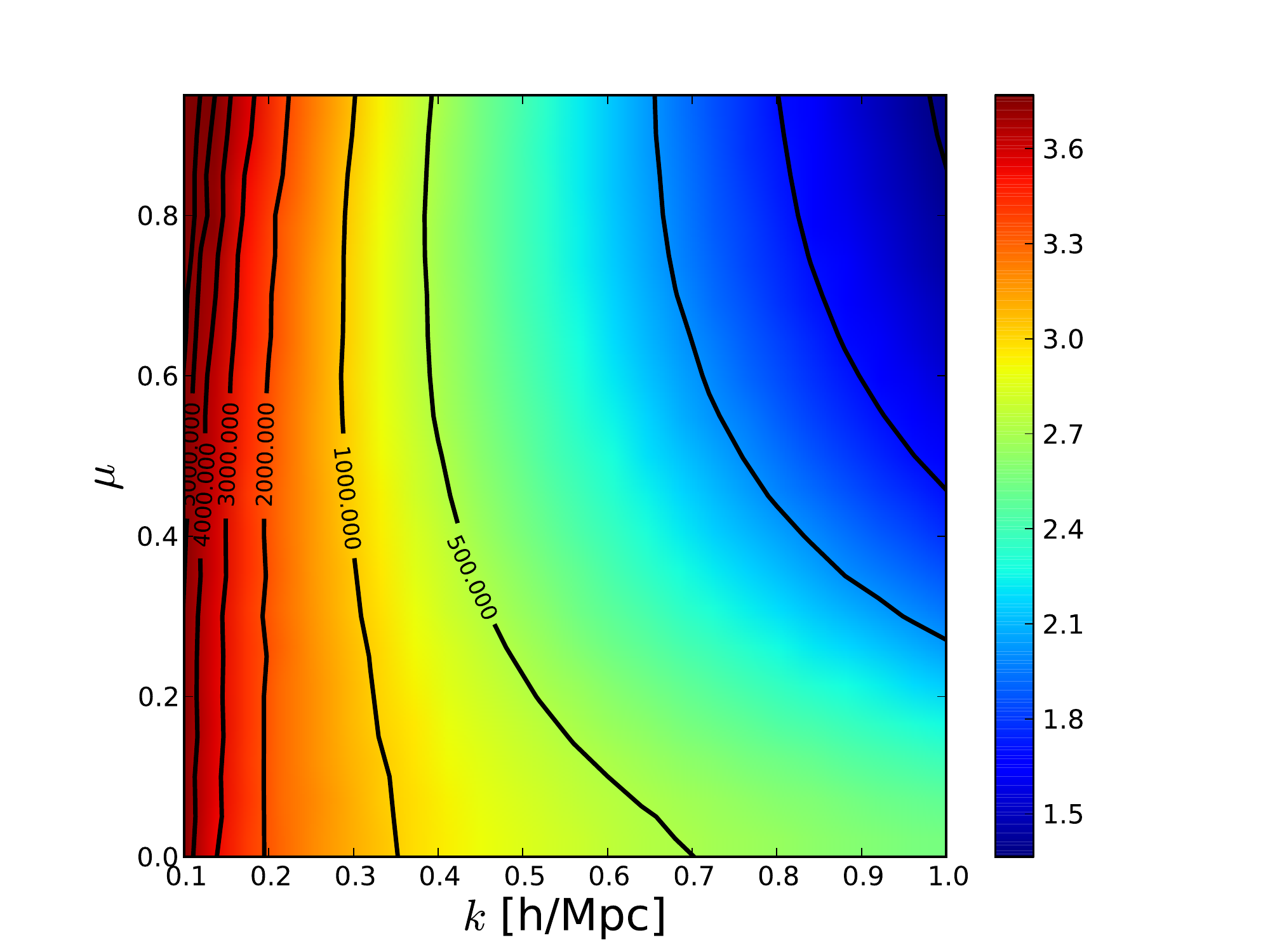}\\ 
\includegraphics[width=\columnwidth]{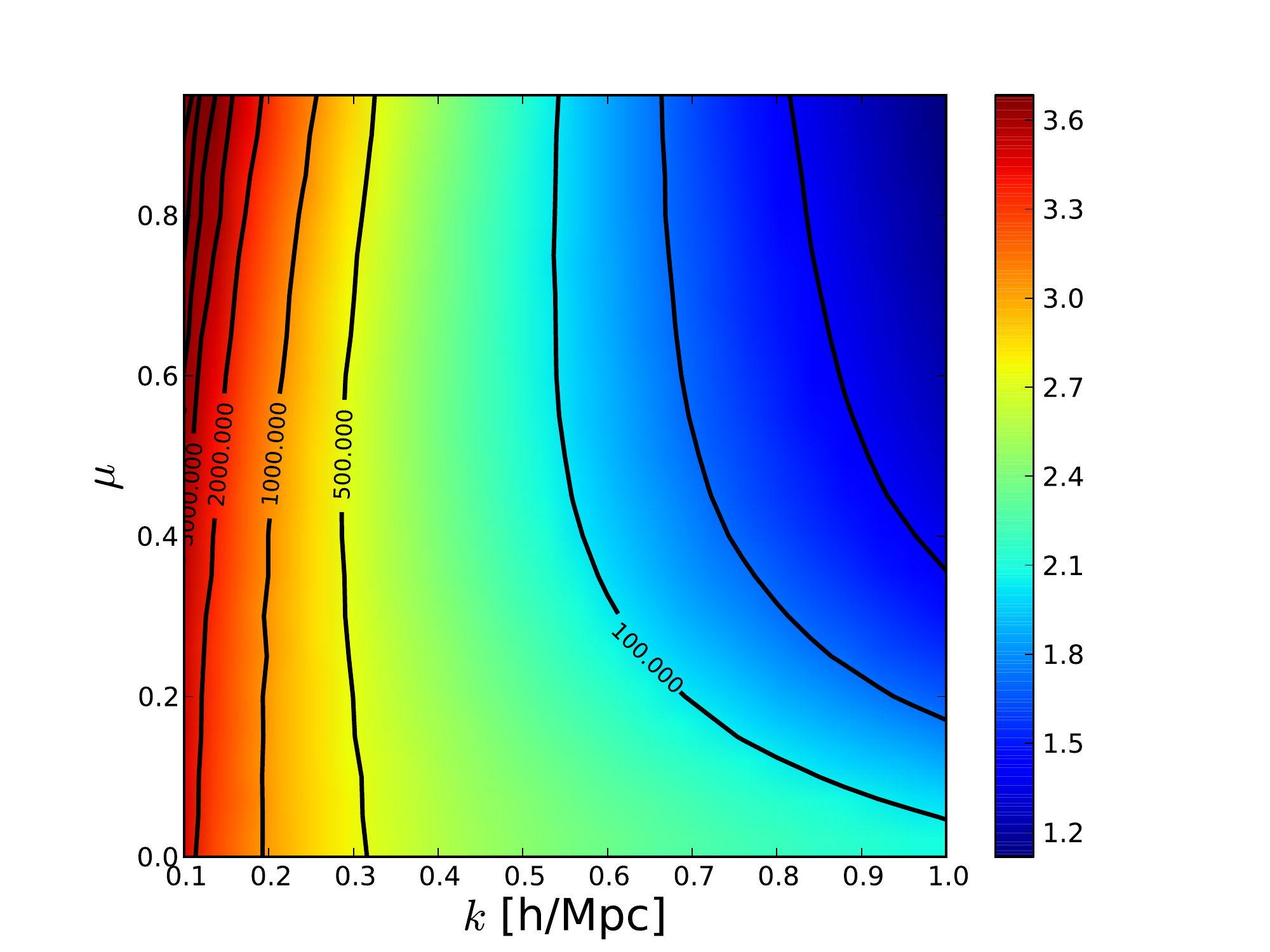}\\ 
\includegraphics[width=\columnwidth]{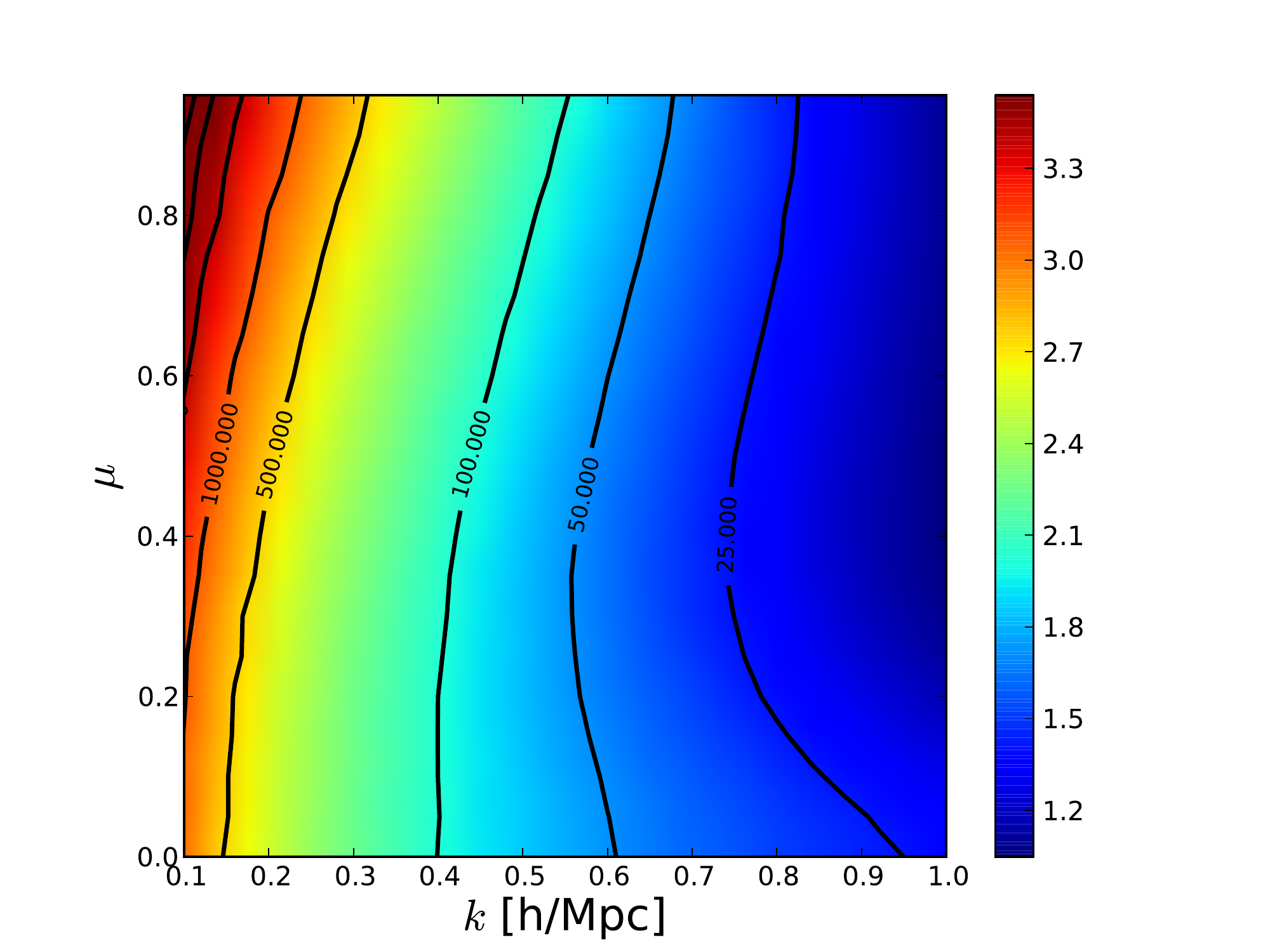}  
\caption{Isocontours of the nonlinear, redshift space power spectrum 
measured on the M000 simulation are plotted in the $k$-$\mu$ plane.  
The color bar shows the log of the power spectrum. 
} 
\label{fig:purep} 
\end{figure}

Figure~\ref{fig:RSD_PS_fractional_difference_M000} shows the fractional 
difference between the simulation measurement and the full, calibrated KLL 
result.  
The calibrated reconstruction is accurate at the 2\% level at low redshift 
and high redshift, with limited regions of the $k$-$\mu$ plane having 
$\gtrsim5\%$ deviations around $z\approx1$.  This appears to be due to the 
interplay of velocity and nonlinear density effects being comparable 
at that epoch.  Overall, the reconstruction of the simulated nonlinear 
redshift space power spectrum shows that the KLL form provides a good 
description over this range of $k$ and $z$.   

Given the simple behaviors of the coefficients seen in 
Fig.~\ref{fig:ABC_M000}, we can explore purely analytic fits for the 
reconstruction function.  
For full calculations one should use the calibrated coefficients, 
but for quick if approximate results it is of interest to see how well 
the analytic approximations perform.  We adopt 
\be 
A=\left[1+\left(\frac{k}{k_A}\right)^{n}\,\right]^{1/n} \ , \label{eq:anlya} 
\ee 
where this form gives a power law increase above 1 at low $k$ and then 
matches at higher $k$ the linear dependence on $k$ observed in the simulation 
results.  Table~\ref{tab:values} gives the values for $k_A$ and $n$ 
at each scale factor, but these approximately follow 
\bea 
k_A(a)&\approx&0.194+(1-a)^{2.4}\\ 
n(a)&\approx&3.2-2.54(1-a) \ . 
\eea 

For $B$, the $k$ dependence is linear over a wide range, so we adopt 
\be 
B=B_0+B_1 k \ , \label{eq:anlyb} 
\ee 
and give the values for $B_0(a)$ and $B_1(a)$ in Table~\ref{tab:values}. 
Similarly for $C$ the exponential $k$ dependence is a good approximation, 
with 
\be
C=C_0\,e^{-k/k_C} \ , \label{eq:anlyc} 
\ee 
and the values for $C_0(a)$ and $k_C(a)$ are in Table~\ref{tab:values}.

\begin{table}[!htb]
\begin{tabular}{l|cccccc} 
 $a$ & $k_A$ & $n$ & $B_0$ & $B_1$ & $C_0$ & $k_C$ \\ 
\hline 
0.333$\ $ & $\ $ 0.75 $\ $& $\ 1.47\ $ & $\ 6.09\ $ & $\ 9.47\ $ & $\ 46.0\ $ & $\ 6.34\ $ \\ 
0.4 & 0.52 & 1.68 & 15.6 & 5.67 & 44.9& 7.07 \\ 
0.5 & 0.37 & 1.94 & 25.7 & 9.79 & 37.1 & 7.45 \\ 
0.6 & 0.30 & 2.17 & 32.5 & 10.8 & 29.6 & 7.43 \\ 
0.7 & 0.25 & 2.42 & 32.2 & 21.2 & 26.8 & 7.41 \\ 
0.8 & 0.23 & 2.64 & 31.5 & 19.8 & 24.4 & 7.24 \\ 
0.9 & 0.21 & 2.87 & 29.3 & 16.4 & 22.2 & 6.95 \\ 
1.0 & 0.19 & 3.23 & $\ 26.0\ $ & $\ 13.5\ $ & 19.7 & 6.54 
\end{tabular}
\caption{The parameter values for the redshift space power spectrum fits of 
Eqs.~(\ref{eq:anlya}-\ref{eq:anlyc}) at each scale factor $a$, 
with $k_A$ and $k_C$ in units of 
$h$/Mpc, $B_0$ and $C_0$ in $({\rm Mpc}/h)^2$, $B_1$ in $({\rm Mpc}/h)^3$, 
and $n$ dimensionless. 
}
\label{tab:values}
\end{table}

These expressions capture the $k$ dependences and do well 
quantitatively at reproducing the coefficient values over the range 
$k=0.25-0.5\,h$/Mpc and $z=0-2$.  The true test, however, is how well 
the reconstructed nonlinear redshift space power spectrum matches the 
truth measured from the simulations. 

Figure~\ref{fig:Anly_PS_fractional_difference_M000} shows the 
fractional difference between the simulation measurement and the purely 
analytic KLL result.  
We see that the analytic approximation does nearly as well as the calibrated 
coefficients in the fit range of $k=0.25-0.5\,h$/Mpc.  
Appendix~\ref{apx:realeasy} considers the accuracy for fits to the real space 
power spectrum.

\section{Converse Approach} \label{sec:kfn} 

The KLL form has demonstrated successful reconstruction of redshift space 
distortions by adopting a functional form for the angular dependence $\mu$ 
and fitting for the $k$ dependence.  One might imagine the converse 
approach of adopting a functional form for the $k$ dependence and fitting 
for the $\mu$ dependence.  Remarkably this shows promise as well, and we 
briefly describe this VL method, leaving its development for future work. 

Part of the reason why this is possible is the smoothness of the redshift 
space power spectrum.  Figure~\ref{fig:purep} shows the redshift space 
power spectrum measured from the M000 simulation at three redshifts.  The 
isocontours of the power spectrum are smooth curves in the $k$-$\mu$ 
plane, gradually changing with redshift, so simple functional forms can 
accurately capture their behavior.

The nonlinear real space power spectrum is a $k$-dependent but 
$\mu$-independent amplification of the linear power spectrum; let us assume 
this is determined directly by simulations and concentrate on the redshift 
space distortion effects.  We write these as 
\be 
P(k,\mu,z)=e^{-(Ek^2+Fk+G)}\,P^r(k,z) \ , \label{eq:psmu} 
\ee 
where the VL reconstruction coefficients $E,F,G$ are all functions 
of $\mu$ and $z$.  

\begin{figure}[!htb]
\includegraphics[width=\columnwidth]{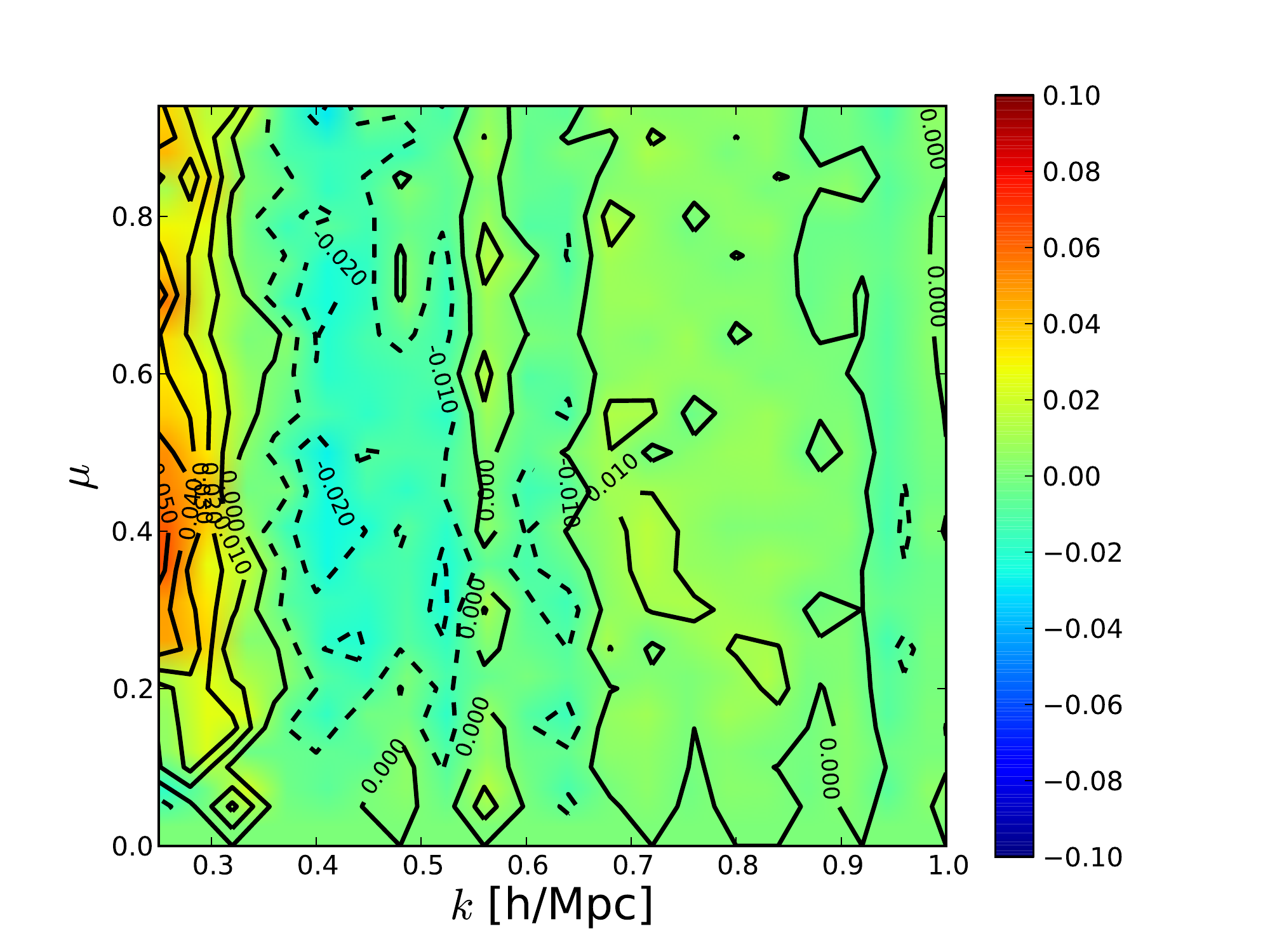}\\ 
\includegraphics[width=\columnwidth]{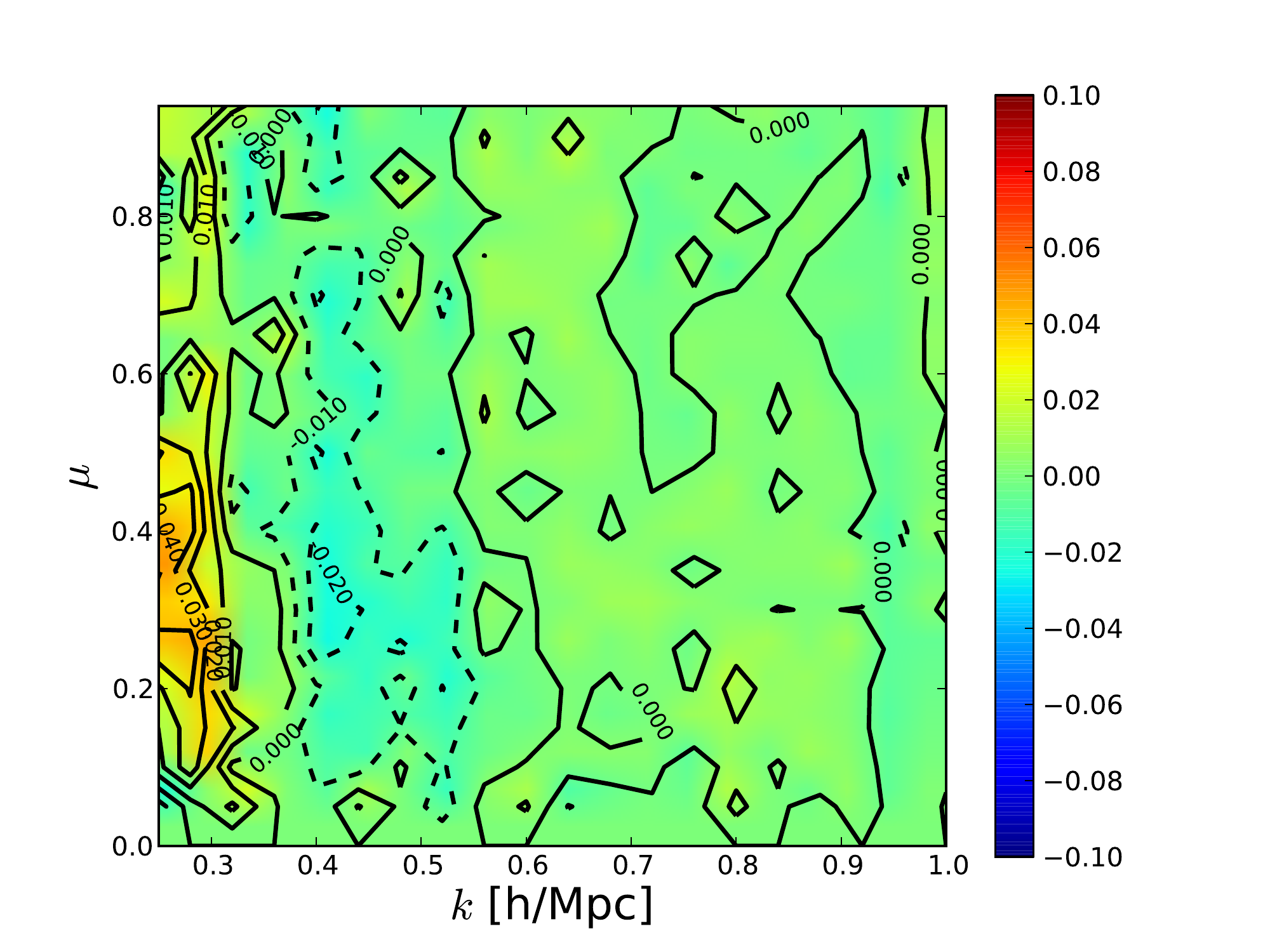}\\ 
\includegraphics[width=\columnwidth]{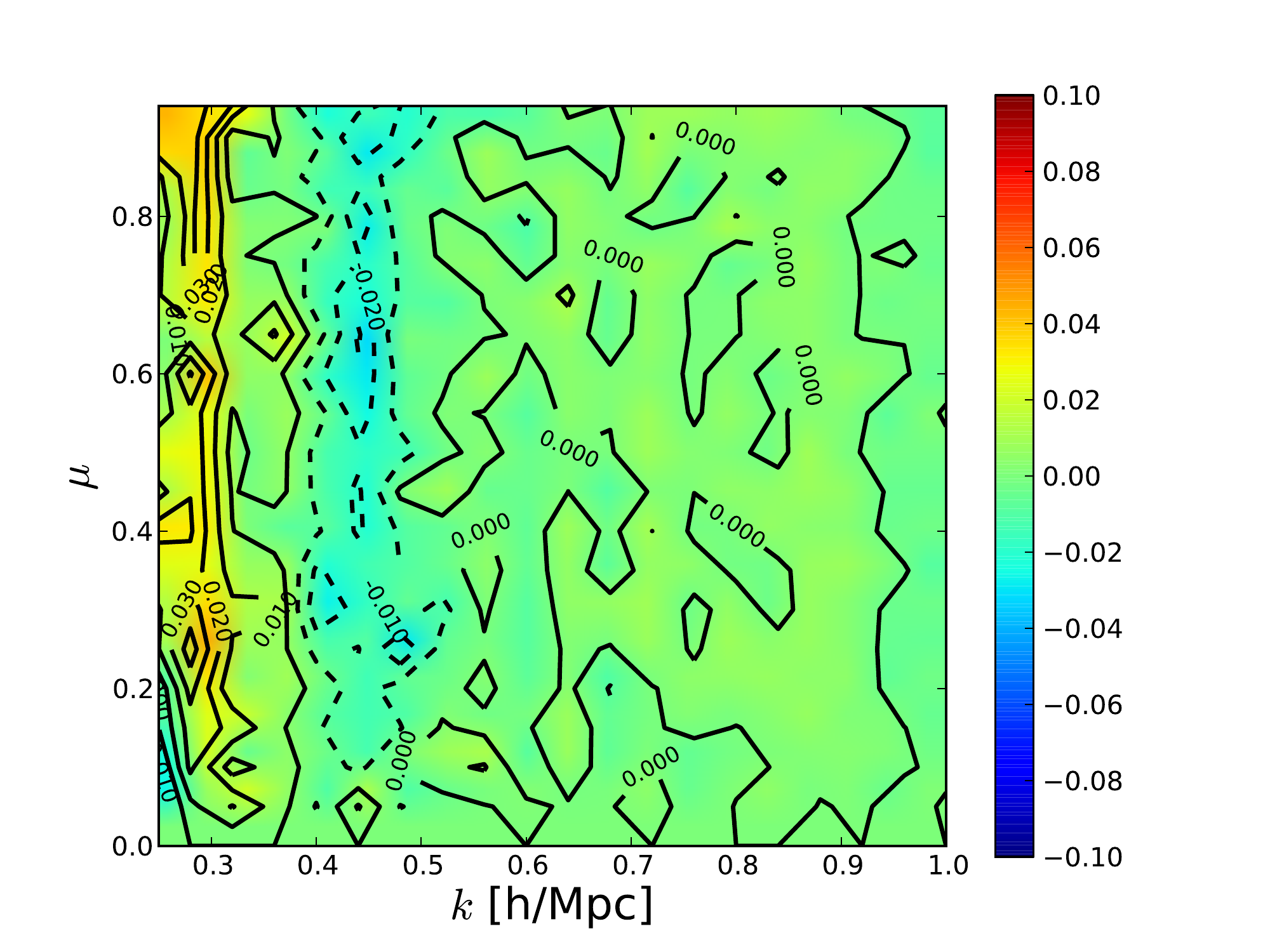} 
\caption{Fractional difference between the RSD power spectrum measured on 
M000 simulation and the one using the simple VL analytic reconstruction from 
Eq.~(\ref{eq:psmu}), at $z=0, 0.67, 2$ (top, center, bottom panel). 
} 
\label{fig:dpsmu} 
\end{figure}

This approach allows general angular dependence, but its $k$ dependence 
is determined by the real space power spectrum and Eq.~(\ref{eq:psmu}). 
Comparison with the redshift space power spectrum measured from the 
simulation shows great promise, with deviations less than 2\% over 
the range $k=0.3-1\,h$/Mpc as seen in Fig.~\ref{fig:dpsmu}.  This 
complementary approach will be the subject of future work.

\section{Conclusions} \label{sec:concl} 

Rich cosmological information resides in the redshift space power spectrum, 
sensitive to the growth rate of cosmic structure.  Connecting the linear, 
isotropic real space power spectrum that is easily predicted from theory 
to the observed, nonlinear, anisotropic redshift space power spectrum is 
a challenge that must be addressed to take advantage of ongoing and future 
spectroscopic surveys.  

One path forward uses the KLL reconstruction function, with a particular 
angular dependence, to carry out the mapping with three coefficients 
depending on wavenumber $k$ and redshift $z$.  Here we have verified that 
this form provides accurate reconstruction over the range $k=0.1-0.5\,h$/Mpc 
and $z=0-2$.  For most of this range the accuracy is at the 2\% level, 
extending significantly beyond the reach of perturbation theory 
and offering hope 
that galaxy survey data can used well beyond $k=0.1\,h$/Mpc with all the 
benefits of the longer leverage and much larger number of Fourier modes. 

Establishing a map between the KLL approach and SMD distribution function 
perturbation expansion, we show how SMD can predict the KLL coefficients 
at low wavenumber, and KLL can then extend beyond SMD's range of validity. 
Direct simulations, however, can calibrate the KLL coefficients efficiently, 
giving the above mentioned 2\% reconstruction accuracy.  

The wavenumber dependence of the coefficients is smooth and simple, and 
even purely analytic approximations prove rather accurate in reconstructing 
the redshift space power spectrum, though we provide a table of optimized 
values of calibrated coefficients instead.  Moreover, the accuracy is 
extremely good for transverse modes -- those undistorted from real space. 
We introduce the RealEasy function for mapping linear to nonlinear power 
spectra in real space, achieving 1\% accuracy, a substantial improvement 
over the commonly used Halofit (albeit for just one model so far). 

We also introduce the converse approach to redshift space distortions, 
allowing arbitrary angular dependence but a form for $k$ dependent 
correction to the nonlinear real space power spectrum.  This VL form 
also shows great promise, with apparently 2\% accuracy for $k=0.3-1\,h$/Mpc.  

Many steps remain in developing practical, robust understanding and use of 
redshift space distortions.  Having established the $k$ and $z$ dependence 
of the reconstruction coefficients, we will next use suites of cosmological 
simulations to investigate their variation with cosmological model.  
The path from dark matter simulations to halos to observed galaxies is 
a nontrivial one and will require strenuous, long term effort.  However, 
the positive indications given by our analysis using the KLL and VL forms, 
backed up the physics of the smoothness of the power spectrum variation in 
the $k$-$\mu$ plane, provide cautious optimism that these approaches 
contain elements useful for improving and extending the information locked 
within spectroscopic surveys.

\acknowledgments 

We sincerely thank Katrin Heitmann, Salman Habib and the Institutional 
Computing program at Los Alamos National Laboratory for kindly allowing 
us access to the Coyote Universe simulation suite.  
We thank Juliana Kwan and Zarija Luki{\'c} for helpful discussions. We 
thank Teppei Okumura for discussions and for sharing his measurements 
of the power spectra $P^{ab}(k)$. 
This work has been supported by DOE grant DE-SC-0007867 and the Director, 
Office of Science, Office of High Energy Physics, 
of the U.S.\ Department of Energy under Contract No.\ DE-AC02-05CH11231, 
and World Class University grant R32-2009-000-10130-0 through the 
National Research Foundation, Ministry of Education, Science and Technology 
of Korea.

\appendix

\section{RealEasy for Real Space} \label{apx:realeasy} 

The residuals of the KLL analytic approximation for the redshift space 
power spectrum are especially small along the $\mu=0$ axis at all redshifts.  
But this is exactly $A(k,z)$, giving the mapping from the 
linear, real space power spectrum $P_L$ to the nonlinear, real space power 
spectrum $P^r$.  While this is tangential to the main exploration of redshift 
space distortions in this article, it is of interest as we can compare 
it directly to the commonly used Halofit mapping \cite{halofit}.  Since 
the $A(k,z)$ in Eq.~(\ref{eq:anlya}) was optimized simultaneously with 
$B$ and $C$, across $\mu$, we can actually obtain a better estimate if we 
concentrate only on $A$, i.e.\ the real space power spectrum. 

Calibrating from the simulation measurements in real space, we find 
the ``RealEasy'' form 
\bea 
P^r(k,z)&=&\left[1+\left(\frac{k}{k_A}\right)^{n}\,\right]^{1/n} P_L(k,z) \label{eq:pr}\\ 
k_A(a)&\approx&0.2+(1-a)^{2.4}\\ 
n(a)&\approx&3.42-2.19(1-a) \ . 
\eea 
The expressions for $k_A$ and $n$ are only approximate, with the actual 
values given in Table~\ref{tab:valuesreal}.

\begin{table}[!htb]
\begin{tabular}{l|cc} 
 $a$ & $k_A$ & $n$\\ 
\hline 
0.333$\ $ & $\ $ 0.56 $\ $& $\ 1.93\ $ \\ 
0.4 & 0.44 & 2.09 \\ 
0.5 & 0.34 & 2.34 \\ 
0.6 & 0.28 & 2.57 \\ 
0.7 & 0.25 & 2.78 \\ 
0.8 & 0.23 & 2.99 \\ 
0.9 & 0.21 & 3.18 \\ 
1.0 & 0.20 & 3.42 
\end{tabular}
\caption{The parameter values for the real space power spectrum fits of 
Eq.~(\ref{eq:pr}) at each scale factor $a$, 
with $k_A$ in units of $h$/Mpc and $n$ dimensionless. 
}
\label{tab:valuesreal}
\end{table}

\begin{figure}[!htb]
\includegraphics[width=\columnwidth]{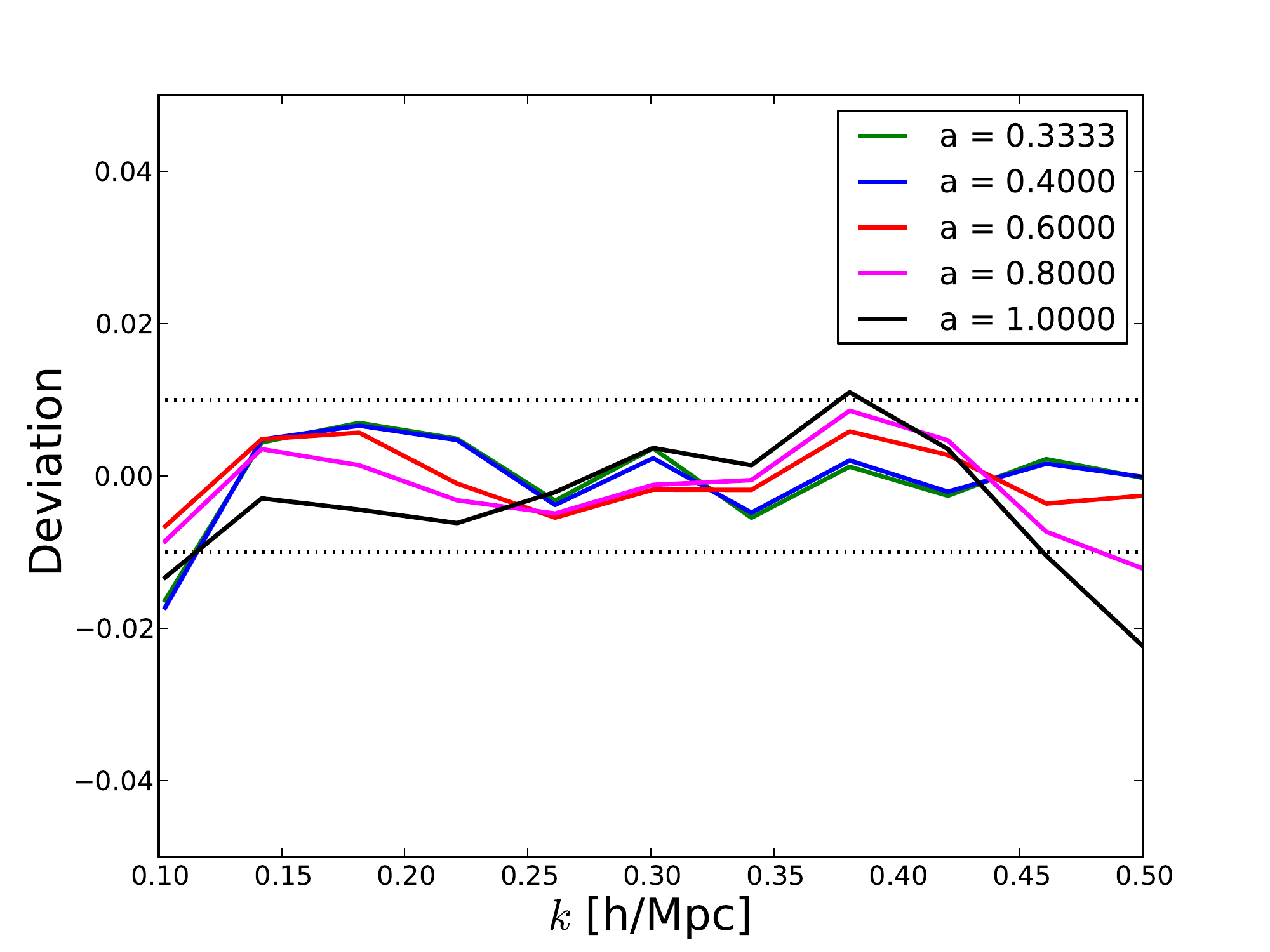} 
\caption{Fractional difference between the nonlinear, real space power 
spectrum measured on M000 and the analytic RealEasy form is plotted vs 
wavenumber.  Curves are for $a=0.33$, 0.4, 0.6, 0.8, 1, from top to bottom 
at $k=0.5\,h$/Mpc. 
} 
\label{fig:anlyfit} 
\end{figure}

Figure~\ref{fig:anlyfit} shows that using Eq.~(\ref{eq:pr}) the 
accuracy in reconstructing the nonlinear, real space 
matter power spectrum is $\sim$1\% over the range 
$k=0.12-0.46\,h$/Mpc.  By comparison, Halofit for \lcdm\ 
is accurate to 5-10\% over this same range \cite{Heitmann:2008eq}, 
with the revised Halofit of \cite{12082701} good to 5\%.  
Moreover, the analytic form of Eq.~(\ref{eq:pr}) is extremely simple.  
We emphasize that so far this has only been tested for the M000 cosmology. 
In the next stage of our work, when we use the other Coyote simulations 
to explore cosmology dependence, it will be interesting to see if the 
form holds and how the parameters depend on cosmology.

\end{document}